\documentclass[12pt]{article}

\usepackage[utf8]{inputenc}
\usepackage{charter, fullpage}
\usepackage{hyperref}
\usepackage{amssymb, upgreek}
\usepackage{graphicx}
\usepackage{hyperref}
\hypersetup{hidelinks}
\usepackage[hybrid]{markdown}
\usepackage[total={6.5in,9in},top=1in,headsep=0.1in,headheight=1in]{geometry}

\usepackage[font={small}]{caption}

\usepackage{lineno}

\newcommand\snowmass{
\begin{center}
  \rule[-0.2in]{\hsize}{0.01in}\\
  \rule{\hsize}{0.01in}\\
  \vskip 0.1in
  Submitted to the Proceedings of the US Community Study\\ 
  on the Future of Particle Physics (Snowmass 2021)\\
 
  \rule{\hsize}{0.01in}\\
  \rule[+0.2in]{\hsize}{0.01in}\\[-2em]
\end{center}
}
\usepackage[firstpage=true]{background}
\backgroundsetup{contents={\parbox{6.5in}{\snowmass}}, scale=1,placement=top,opacity=1,color=black,position={3.25in,1.2in}}

\usepackage{fancyhdr}
\fancypagestyle{plain}{%
  \fancyhf{}%
  \fancyhead[C]{}
  \fancyfoot[C]{\thepage}
}

\fancypagestyle{empty}{%
  \fancyhf{}%
  \fancyhead[C]{{\it Liquid-noble Bubble Chambers for Dark Matter and CE$\nu$NS Detection}}
  \fancyfoot[C]{\thepage}
}
\title{Scintillating Bubble Chambers: \\ Liquid-noble Bubble Chambers for Dark Matter and CE$\nu$NS Detection}
\date{}

\usepackage{authblk}

\author[1]{E.~Alfonso-Pita}
\author[2]{M.~Baker}
\author[3]{E.~Behnke}
\author[4]{A.~Brandon\thanks{now at Northrop Grumman}}
\author[5]{M.~Bressler\thanks{now at UMass Amherst}}
\author[6]{B.~Broerman}
\author[6]{K.~Clark\thanks{kenneth.clark@queensu.ca}}
\author[4]{R.~Coppejans\thanks{now at Univeristy of Oxford}}
\author[6]{J.~Corbett}
\author[3]{C.~Cripe}
\author[7]{M.~Crisler}
\author[4,7]{C.E.~Dahl\thanks{cdahl@northwestern.edu}}
\author[6]{K.~Dering}
\author[6]{A.~de St.~Croix}
\author[2]{D.~Durnford}
\author[6]{K.~Foy}
\author[8]{P.~Giampa}
\author[4]{J.~Gresl\textsuperscript{*}}
\author[8]{J.~Hall}
\author[9]{O.~Harris}
\author[6]{H.~Hawley-Herrera}
\author[10]{C.M.~Jackson}
\author[4]{M.~Khatri}
\author[2]{Y.~Ko}
\author[5]{N.~Lamb}
\author[11]{M.~Laurin}
\author[3]{I.~Levine}
\author[12]{W.H.~Lippincott}
\author[4]{X.~Liu}
\author[5]{R.~Neilson}
\author[2]{S.~Pal}
\author[4]{J.~Phelan\thanks{now at MIT}}
\author[2]{M.-C.~Piro}
\author[13]{S.~Priya}
\author[4]{S.~Ray}
\author[4]{E.~Rich}
\author[4]{Z.~Sheng}
\author[4]{A.~Sloss}
\author[4]{X.~Struyk}
\author[1]{E.~V\'azquez-J\'auregui}
\author[4]{D.~Velasco}
\author[14]{S.~Westerdale}
\author[12]{T.J.~Whitis}
\author[4]{W.~Zha}
\author[12]{R.~Zhang}

\affil[1]{Instituto de F\'isica, Universidad Nacional Autónoma de M\'exico, A.P. 20-364, Ciudad de M\'exico 01000, M\'exico.}
\affil[2]{Department of Physics, University of Alberta, Edmonton, T6G 2E1, Canada}
\affil[3]{Department of Physics and Astronomy, Indiana University South Bend, South Bend, Indiana 46634, USA}
\affil[4]{Department of Physics and Astronomy, Northwestern University, Evanston, Illinois 60208, USA}
\affil[5]{Department of Physics, Drexel University, Philadelphia, Pennsylvania 19104, USA}
\affil[6]{Department of Physics, Engineering Physics, and Astronomy, Queen’s University, Kingston, K7L 3N6, Canada}
\affil[7]{Fermi National Accelerator Laboratory, Batavia, Illinois 60510, USA}
\affil[8]{SNOLAB, Lively, Ontario, P3Y 1N2, Canada}
\affil[9]{Northeastern Illinois University, Chicago, Illinois 60625, USA}
\affil[10]{Pacific Northwest National Laboratory, Richland, Washington 99354, USA}
\affil[11]{D\'epartement de Physique, Universit\'e de Montr\'eal, Montr\'eal, H3T 1J4, Canada}
\affil[12]{Department of Physics, University of California Santa Barbara, Santa Barbara, California 93106, USA}
\affil[13]{Materials Research Institute, Pennsylvania State University, University Park, Pennsylvania 16802, USA}
\affil[14]{Department of Physics and Astronomy, University of California Irvine, Irvine, California 92697, USA}

\begin{document}


\maketitle
\begin{abstract}

The Scintillating Bubble Chamber (SBC) collaboration is developing liquid-noble bubble chambers for the quasi-background-free detection of low-mass (GeV-scale) dark matter and coherent scattering of low-energy (MeV-scale) neutrinos (CE$\nu$NS).  The first physics-scale demonstrator of this technique, a 10-kg liquid argon bubble chamber dubbed SBC-LAr10, is now being commissioned at Fermilab.  This device will calibrate the background discrimination power and sensitivity of superheated argon to nuclear recoils at energies down to 100 eV.  A second functionally-identical detector with a focus on radiopure construction is being built for SBC's first dark matter search at SNOLAB.  The projected spin-independent sensitivity of this search is approximately $10^{-43}$ cm$^2$ at 1~GeV$/c^2$ dark matter particle mass.  The scalability and background discrimination power of the liquid-noble bubble chamber make this technique a compelling candidate for future dark matter searches to the solar neutrino fog at 1~GeV$/c^2$ particle mass (requiring a $\sim$ton-year exposure with non-neutrino backgrounds sub-dominant to the solar CE$\nu$NS signal) and for high-statistics CE$\nu$NS studies at nuclear reactors.

\end{abstract}
\newpage
\section*{Executive Summary}\label{sec:execsummary}
The Scintillating Bubble Chamber (SBC) collaboration is developing liquid-noble bubble chambers for large (ton-yr) exposure, quasi-background-free detection of low-energy (sub-keV) nuclear recoils.  This effort addresses the current lack of a scalable, background-discriminating, low-threshold nuclear recoil detection technology, enabling searches for low-mass ($\sim$1~GeV$/c^2$) dark matter to the neutrino fog and high-statistics measurements of low-energy ($\mathcal{O}(1)$~MeV) neutrino coherent scattering.  

The liquid-noble bubble chamber combines the field-leading electron-recoil background discrimination of a bubble chamber with the event-by-event energy resolution of a liquid scintillator detector.  Early demonstrations of this technique revealed that competition between these two detection channels, and more generally the lack of bubble-nucleating energy-loss mechanisms available to recoiling electrons in liquid nobles, leads to much \emph{stronger} electron-recoil discrimination than in existing Freon-based bubble chamber dark matter detectors.  This in turn allows background-free sensitivity to nuclear recoils at lower energies than can be achieved in any currently working detection technique, potentially reaching nuclear recoil detection thresholds as low as 100~eV \emph{while remaining blind to electron-recoil backgrounds.}  This feature, together with the inherent scalability of liquid-based detectors, removes the critical obstacle to both GeV-scale dark matter detection and coherent neutrino scattering (CE$\nu$NS) measurements at nuclear reactors.

The low-threshold performance of the liquid-noble bubble chamber technique will be validated in the SBC collaboration's first physics-scale device, the 10-kg liquid argon bubble chamber  ``SBC-LAr10.''  This device, now being commissioned, will undergo a series of source calibrations in the MINOS underground facility at Fermilab to measure both the discrimination power and low-energy nuclear recoil sensitivity of the device, calibrating the nuclear recoil detection threshold with a resolution of 10~eV.  At the same time, a radiopure, functionally-identical clone of SBC-LAr10 is being constructed for the SBC collaboration's first dark matter search at SNOLAB, with a projected sensitivity to spin-independent dark matter-nucleon scattering of $2\cdot10^{-43}$---$10^{-44}$~cm$^2$ in the 1---10~GeV$/c^2$ mass range.

Future SBC efforts will include a ton-year dark matter search targeting the solar-neutrino fog at 1~GeV$/c^2$ particle mass and the deployment of an SBC device at a nuclear reactor to measure the CE$\nu$NS cross section at $\mathcal{O}(1)$~MeV neutrino energy. The technical challenges in these efforts are modest, taking advantage of expertise gained by SBC collaborators from all corners of the low-background field, including large-scale bubble chamber operation (\emph{e.g.} PICO), cryogenics and liquid-noble handling (\emph{e.g.} LUX-ZEPLIN), low-threshold calibration (\emph{e.g.} NEWS-G), and low-background construction and photon-detection techniques (\emph{e.g.} nEXO).

Section~\ref{sec:science} of this white paper describes the unique dark matter and neutrino physics opportunities targeted by the SBC collaboration; Section~\ref{sec:concept} reviews the fundamental principles of the SBC technique, supported by data in bench-scale prototypes;  Section~\ref{sec:design} presents the technical design of the SBC-LAr10 device;  Section~\ref{sec:calibration} describes the ongoing effort to  calibrate the sensitivity and discrimination power of the liquid-noble bubble chamber technique; and Section~\ref{sec:future} outlines the SBC collaboration's longer term plans to fulfill the physics mission described in Section~\ref{sec:science}.


\newpage
\section{Science Opportunities}\label{sec:science}
Measurements of the cosmic abundances of both baryonic and dark matter are among the greatest triumphs of modern cosmology and are a cornerstone of our understanding of the universe, but they also give rise to the two greatest open questions in high energy physics:  {\bf What is the dark matter?} and {\bf What drove the matter/anti-matter asymmetry in the early universe?}  The liquid-noble bubble chamber is an enabling technology on both of these fronts, the former through dark matter direct detection searches and the latter through precision measurements of neutrino-nucleus cross-sections, where hints may appear of new interactions linking the asymmetry of the neutrino sector to that observed cosmically in baryons.

The desired signal in a liquid-noble bubble chamber, for both dark matter and neutrino interactions, is the elastic scattering of low-momentum particles off atomic nuclei.  With a target detection threshold of 3~MeV$/c$ in momentum transfer (\emph{e.g.}, producing a 100-eV recoil on an argon nucleus) this technique can search for dark matter particles with masses down to $\sim$1~GeV$/c^2$ and observe coherent elastic neutrino-nucleus scattering (CE$\nu$NS) from neutrinos and anti-neutrinos with energies down to 1.5~MeV, just below the 1.8-MeV hydrogen inverse-beta-decay threshold.

Unlike other sub-keV nuclear recoil detection techniques, the liquid-noble bubble chamber is both scalable and ``quasi-background-free,'' meaning that the technique is able to discriminate against all backgrounds to the dark matter and CE$\nu$NS signals at the level needed to reach expected backgrounds of $<$1 event in the signal region of interest over an experiment's lifetime.  We expect quasi-background-free exposures of $\mathcal{O}(1)$~ton-year to be achievable in a typical low-background (underground, shielded) environment, noting that at these exposures the solar neutrino CE$\nu$NS and potential dark matter signals become backgrounds to each other, forming the so-called ``neutrino fog''~\cite{OHare:2021utq}.  As in high-mass dark matter searches, liquid-noble bubble chambers achieve background-free exposures through a combination of (1) applying a narrow energy region of interest, (2) identifying multiply-scattering backgrounds through a combination of position reconstruction within the target and active vetoes surrounding the target, and (3) differentiating between electron-recoil (ER) backgrounds (\emph{e.g.}, from beta-decay) and the nuclear-recoil (NR) signal.  No technology has yet shown ER/NR discrimination below 1~keV (NR energy), and of the technologies with the \emph{potential} for such discrimination, only liquid-noble bubble chambers have the scalability needed to reach the ton-year exposures motivated by the dark matter and neutrino physics goals described below.

\subsection{Dark Matter Detection at 1~GeV/$c^2$}

Many dark matter candidates, including Weakly Interacting Massive Particles (WIMPs) and dark-sector models, populate the 1--10~GeV$/c^2$ mass range~\cite{Akerib:2022ort}.  
This region is of particular interest for asymmetric dark matter (ADM) models~\cite{Cohen:2010kn}, where the similar cosmic abundances of dark and baryonic matter are not a coincidence but rather a consequence of a coupled matter/anti-matter asymmetry in the light and dark sectors. ADM models can span a wide range of masses, but naturally favor dark matter particles at a few times the proton mass.  These models are inaccessible to indirect detection as they leave too little anti-dark-matter in the present-day universe to create an annihilation signal, and at 1~GeV$/c^2$ are also above the mass range probed by beam-dump experiments~\cite{LDMX:2018cma}.  This leaves direct detection as the sole method to explore this key mass range, and was the primary motivation for one third of the ``Generation~2'' U.S. dark matter search program~\cite{Agnese:2016cpb}.

\begin{figure}[!bth]
    \centering
    \includegraphics[height=2.6in, trim = 0 0 0 56, clip=true]{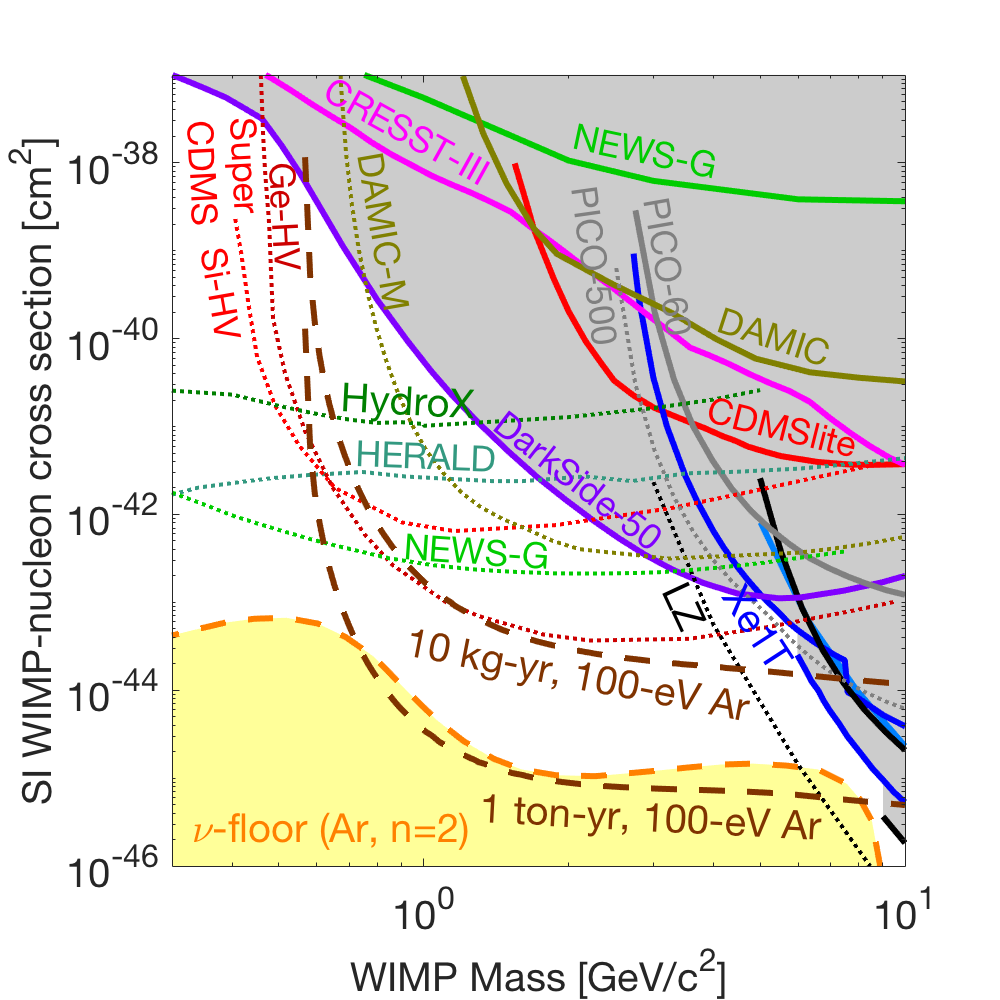}
    \includegraphics[height=2.6in, trim = 0 0 0 0, clip=true]{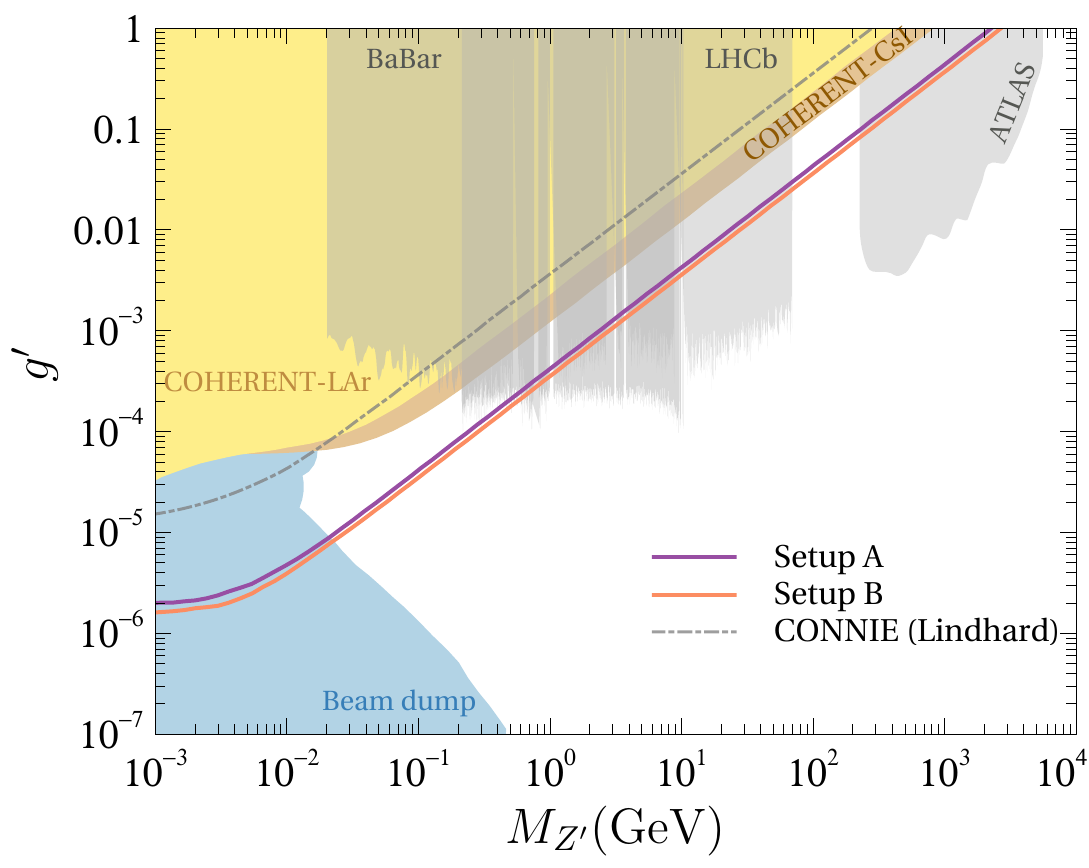}
    \caption{{\bf Left:}  Current limits and projections in the search for GeV-scale dark matter.  All curves are 90\% C.L. limits (solid lines~\cite{Aalbers:2022fxq,Aprile:2019xxb,Agnes:2018fg,Aprile:2018dbl,PhysRevLett.118.021303,PhysRevLett.119.181302,Agnese:2018gze,Arnaud:2018a,Aguilar-Arevalo:2020oii,Abdelhameed:2019hmk,Amole:2019fdf,DarkSide:2022dhx}) or projected limits in the absence of a dark matter signal (dotted lines~~\cite{Agnese:2016cpb,Hertel:2018aal,Akerib:2018lyp,settimo2020search,Monte:2019a}) on the spin-independent DM-nucleon scattering cross section.  Additional projections circulated during the Snowmass process are collected at~\cite{Akerib:2022ort}.
    The dashed brown curves are benchmarks indicating the exposure and threshold necessary to reach the solar CE$\nu$NS floor, assuming no background except for the CE$\nu$NS signal, which is subtracted assuming zero uncertainty on the CE$\nu$NS rate.   The neutrino (CE$\nu$NS) floor is taken from~\cite{OHare:2021utq}, where ``$n=2$'' indicates the point at which sensitivity grows with the square-root of exposure, or roughly where systematic uncertainties in the CE$\nu$NS signal overtake statistical uncertainty in the measured CE$\nu$NS rate.
     {\bf Right:} 
     Sensitivity to a new B$-$L coupling for an argon nuclear-recoil detector with 100-eV threshold measuring coherent scattering rates of reactor neutrinos, showing exclusion limits at 95\% C.L. in the $g'-M_{Z'}$ plane.  Setup A refers to a 10-kg chamber 3\,m from the 1-MW$_{\rm{th}}$ research reactor at ININ near Mexico City.  Setup B refers to a 100-kg chamber 30\,m from a 2-GW$_{\rm{th}}$ power reactor.  Both curves include background models (excluding electron-recoil backgrounds) and neutrino flux uncertainty.  Current limits from the CONNIE reactor CE$\nu$NS search \cite{Aguilar-Arevalo:2019zme}, stopped-pion-decay CE$\nu$NS measurements \cite{Akimov:2017ade,Akimov:2020pdx,Akimov:2020czh}, collider experiments \cite{Lees:2014xha,Aaij:2019bvg,Aad:2019fac}, and beam dump experiments \cite{Bergsma:1985qz,Bernardi:1985ny,Konaka:1986cb,Riordan:1987aw,Bjorken:1988as,Davier:1989wz,PhysRevLett.67.2942,Astier:2001ck,Tsai:2019mtm,Banerjee:2019hmi} are also shown.  Plot published in~\cite{SBC:2021yal}.
     }
    \label{fig:sbc_chamber_reach}
\end{figure}

Figure~\ref{fig:sbc_chamber_reach} (left) shows the current state of the dark matter direct detection field in the 1--10~GeV$/c^2$ mass range.  This mass range is below the threshold of the ER/NR-discriminating experiments and analyses employed in heavy WIMP searches (\emph{e.g.}, the lowest mass considered by the recent LUX-ZEPLIN result is 9~GeV$/c^2$~\cite{Aalbers:2022fxq}).  Instead, existing direct detection experiments searching for dark matter at masses below 10~GeV$/c^2$ do so exclusively by measuring very small amounts of ionization (few-$e^-$), either with low-noise electronic readout techniques~\cite{SENSEI:2020dpa} or through built-in gain mechanisms such as proportional gain~\cite{NEWS-G:2022kon},  electroluminescence~\cite{LZ:2019sgr,Aprile:2019xxb,Agnes:2018fg,DarkSide-50:2022qzh}, or Luke-Neganov phonon production~\cite{SuperCDMS:2017nns,SuperCDMS:2018mne}.  This approach has two problems.  First, it measures only a fraction of the available energy:  in the case of electron recoil detection the kinematics of dark matter-electron scattering limit the allowed energy transfer, and in the case of nuclear recoil detection only a small fraction (in some cases $<$10$\%$~\cite{SuperCDMS:2022nlc}) of the transferred energy goes towards ionization due to the Lindhard effect~\cite{Lindhard:1963a}.  Second, with only a single sensitive energy deposition channel it is impossible to distinguish electron- and nuclear-recoil signals, putting quasi-background-free operation well out of reach for both current and next-generation experiments.  {\bf All ionization-based GeV-scale dark matter searches are fundamentally background-limited,} depending critically on an experiment's ability to model and subtract sub-keV ER backgrounds.  To reach the solar CE$\nu$NS fog, and particularly to make a discovery of GeV-scale dark matter, a new low-energy ER/NR discriminating technique is needed. 

Including liquid-noble bubble chambers, there are three technologies that show potential for ER/NR discrimination below 1~keV. The detector physics driving this capability in liquid-noble bubble chambers is the subject of Section~\ref{sec:concept} --- the other two techniques are ultra-high-resolution cryogenic ($<$100~mK) calorimeters measuring both direct quasi-particle generation and electronic energy loss in either germanium/silicon crystals~\cite{Hong:2019zlm} or superfluid helium~\cite{Hertel:2018aal}.  It is reasonable to suppose that both of these techniques will eventually demonstrate the low-energy ER/NR discrimination needed to suppress ER backgrounds below the $\mathcal{O}$(100)~(ton-yr-keV)$^{-1}$ solar neutrino CE$\nu$NS rate, but it will be exceedingly challenging for either to reach the ton-year exposures needed to explore the solar neutrino fog.  Bubble chamber dark matter experiments targeting ton-year exposures, on the other hand, are already going forward for high-mass WIMP searches~\cite{Giroux:2021vpy}. Furthermore, the Excess working group~\cite{adari2022excess} has recently identified a possible explanation for low-energy backgrounds seen in several dark matter search experiments as being due to the release of stored mechanical energy from stresses induced by detector holders or microfractures, or from stress intrinsic to target materials. Liquid-target detectors do not suffer from these types of backgrounds, making liquid-noble bubble chambers a critical element to any future GeV-scale dark matter discovery.

If quasi-background-free operation in a liquid-argon bubble chamber at 100-eV threshold is achieved, a ton-year exposure will explore the solar CE$\nu$NS neutrino fog at 1~GeV$/c^2$ to the $n=2$ boundary as defined in \cite{OHare:2021utq}, or roughly to the point where systematic uncertainties in the solar neutrino model exceed the statistical uncertainty in the measured scattering rate.  In other words, at a ton-year exposure, SBC shifts from performing a dark matter search to exploring solar neutrino physics through the CE$\nu$NS channel. This complementarity has been explored in~\cite{de2021complementarity}. 

\subsection{Coherent Neutrino Scattering at $\mathcal{O}$(1~MeV)}

The CE$\nu$NS interaction was first observed by the COHERENT collaboration, measuring neutrinos produced by stopped-pion and -muon decay at the Oak Ridge National Laboratory's Spallation Neutron Source (SNS) scattering off cesium and iodine nuclei, producing recoils with momenta of $\sim$50~MeV$/c$ ($\sim$10~keV recoil energy)~\cite{Akimov:2017ade}.  COHERENT has since measured this process on argon nuclei as well~\cite{Akimov:2020pdx}, but the CE$\nu$NS process from either solar neutrinos or nuclear reactor anti-neutrinos has yet to be measured. 

The neutrino physics motivation for a reactor CE$\nu$NS measurement is driven by three qualitative features.  First, reactor neutrinos are an order of magnitude lower in energy than the neutrinos produced by a stopped-pion source, producing fully-coherent scattering and probing BSM neutrino physics in a new regime --- \emph{i.e.}, the neutrino-nucleus effective field theory at $\sim$2~MeV contains different information than the effective field theory at $\sim$50~MeV.  Second, the neutrino flux at a typical reactor site is up to $10^{5}$ times higher than that at the SNS.  Finally, reactors are a pure $\bar\nu_e$ source, complementing the mixed $\bar\nu_e, \nu_e, \bar\nu_\mu, \nu_\mu$ produced at spallation sources.  Figure~\ref{fig:sbc_chamber_reach} (right) shows a case study of the physics reach of a reactor CE$\nu$NS measurement using an SBC chamber~\cite{SBC:2021yal}, examining sensitivity to a new B--L coupling; other specific physics targets include other non-standard interactions~\cite{Barranco:2007tz,Dutta:2015vwa}, sterile neutrino oscillations at short-baselines~\cite{Kosmas:2017zbh, Dutta:2015nlo} and neutrino magnetic moments~\cite{Kosmas:2015vsa}. Additional studies on the reach of SBC to non-standard neutrino interactions and sterile neutrinos have been performed~\cite{alfonso2022new}. A precision measurement can also be made of the weak mixing angle at low momentum transfer~\cite{Canas:2018rng, Kosmas:2015vsa}.

There are currently operating or planned efforts to observe reactor CE$\nu$NS with a number of techniques including p-type point contact Ge (CONUS, TEXONO), Si CCD (CONNIE), NaI (NEOS), dual-phase Xe TPC (RED-100, NUXE), cryogenic Si, Ge and Zn bolometer (MINER, RICOCHET) and cryogenic CaWO$_4$ and Al$_2$O$_3$ calorimeter (NUCLEUS) detectors~\cite{Bonet:2020awv,WONG2010229c,Aguilar-Arevalo:2019zme,Atif:2020glb,Akimov:2019ogx,Wei:2020cwl,Agnolet:2016zir,Billard:2016giu,Strauss:2020ujc}. 
Recent results reported by CONUS~\cite{Bonet:2020awv} elucidate the challenge faced by these efforts; due to high background and limited reactor-off time available for background subtraction, the reported upper limit remains a factor of seven above the Standard Model CE$\nu$NS prediction. As in GeV-scale dark matter searches, sub-keV electron-recoil discrimination is the most effective means of background reduction, granting SBC the potential to achieve a signal-to-background ratio much better than one. 

In addition to searches for new neutrino physics, neutrino detectors have been considered for several applications related to nuclear non-proliferation~\cite{Bernstein:2019hix, cogswell2022cerium}. In particular, LAr detectors sensitive to CE$\nu$NS may have applications in near-field, non-intrusive, real-time reactor power monitoring~\cite{Akimov:2009ht}.
 A 10-kg-scale SBC detector has the potential to make a significant CE$\nu$NS detection of neutrinos from a nuclear reactor in much less than a day.

\section{Scintillating Bubble Chambers}\label{sec:concept}
Bubble chambers for dark matter searches, liquid-noble or otherwise, take advantage the field's most straightforward and powerful ER/NR discrimination technique:  nuclear recoils make bubbles in the superheated liquid target and electron recoils do not.  Underlying this statement is a trade-off between the nuclear recoil detection threshold and the electron recoil discrimination power, both of which are determined by the thermodynamic state (temperature and pressure) of the target fluid.  The key phenomenon driving the SBC program is a fundamental, additional suppression of ER-induced bubble nucleation in superheated noble liquids, enabling quasi-background-free operation at NR detection thresholds an order of magnitude or more below those achievable in existing Freon-based chambers.

This section gives a brief review of bubble chamber operation and the unique low-threshold capability of liquid-noble bubble chambers.  For an in-depth modern review of the thermodynamic theory governing bubble chamber operation, see the appendices of~\cite{Amole:2019scf}.  For comparisons between the thermodynamic model and experimental data in superheated molecular fluids, see~\cite{Amole:2019scf,Bressler:2021ubm} (electron-recoil calibrations) and \cite{PICO:2022nyi} (nuclear-recoil calibrations).  For experimental data from the SBC liquid-xenon prototypes, see~\cite{Baxter:2017ozv} (first data) and \cite{BresslerPhD} (low-threshold data).

\subsection{Bubble Chamber Operation}
\label{sec:BCoperation}
The scintillating bubble chamber is an extension of the moderately-superheated bubble chamber technique pioneered by COUPP \cite{Behnke:2008zza} and now used by PICO \cite{Amole:2019fdf} to search for dark matter.  The target material in any of these devices is a homogeneous fluid that is superheated by dropping the pressure of the fluid below its vapor pressure at a given temperature.  This state is maintained, for $\mathcal{O}$(10) minutes or more in a low-background environment, until a particle interaction creates a nucleation site or ``proto-bubble'' in the fluid.  Within milliseconds the bubble grows to visible size and continues to grow until the chamber is compressed by the experimenter, driving the fluid back to a stable liquid state.  After a roughly half-minute delay to re-equilibrate, the pressure is reduced again to place the liquid back in the superheated state.  We refer to this cycle, from stable liquid, to superheated liquid, to bubble, and back to stable liquid, as an ``event.''

The great advantage of the bubble chamber for dark matter searches is its intrinsic insensitivity to electron-recoil backgrounds.  This insensitivity is a direct consequence of the thermodynamics of bubble nucleation, which require the formation of a critically sized proto-bubble to overcome the free energy barrier arising from the surface tension of the bubble~\cite{Gibbs:1928a}.  
A particle interaction can create this critically-sized proto-bubble by locally heating the fluid, as described in the Seitz ``Hot-Spike'' model~\cite{Seitz:1958nva}.  In typical PICO chambers an energy deposition of a few keV is required to create the necessary $\sim$50-nm-diameter proto-bubble~\cite{Amole:2019fdf}. 
Low-energy nuclear recoils meet this local heating criterion, and the Seitz calculation is a good approximation of the measured nuclear recoil energy threshold for bubble nucleation \cite{Amole:2019fdf,PICO:2022nyi,Behnke:2013sma,Archambault:2010jb,Jin:2019thesis}.
Recoiling electrons, on the other hand, cannot achieve the required $\sim$100-eV/nm stopping power, allowing PICO chambers to set the operating conditions such that only 1 in $10^{10}$ electron recoil events will nucleate a bubble.  A comprehensive picture of ER/NR discrimination in bubble chambers and a complete  derivation of the critical radius and Seitz thermodynamic threshold are given in \cite{Amole:2019scf}.

\subsection{Liquid-Noble Bubble Chambers}
\begin{figure}[!bt]
\centering
\includegraphics[height=4in,trim=0 0 310 0, clip=true]{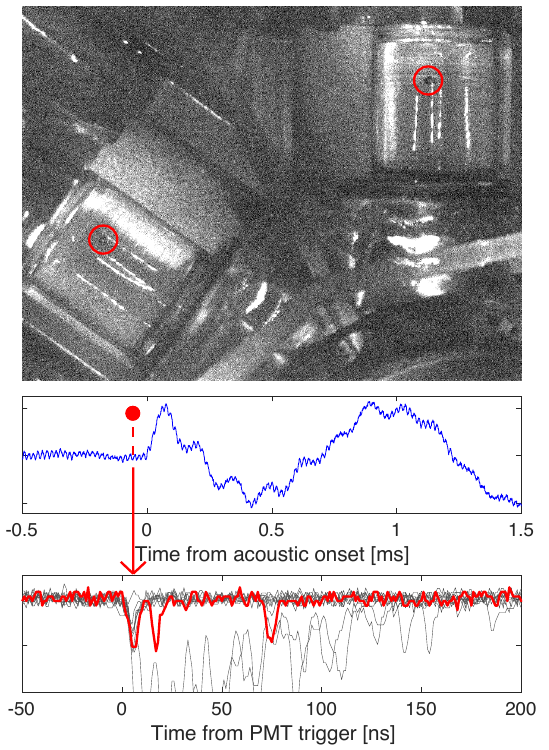}
\caption{
Sample nuclear recoil event from the prototype xenon bubble chamber \cite{Baxter:2017ozv}.  \textbf{Top:}  Stereo image of a single xenon vapor bubble.  \textbf{Middle:} Acoustic record (blue) of bubble formation, giving the time of nucleation to $\pm$25~$\upmu$s.  In this case nucleation is coincident with a scintillation trigger (red). The lag between the scintillation pulse and acoustic onset matches the sound speed in liquid xenon.  \textbf{Bottom:}  PMT waveforms showing xenon scintillation. The bubble-coincident pulse is shown in red.  Electron recoils generate scintillation pulses without coincident bubble nucleation (gray traces).  }
\label{fig:xebc}
\end{figure}

The liquid-noble bubble chambers built by the SBC collaboration differ from the Freon bubble chambers operated by COUPP and PICO in two fundamental, related ways.  The obvious difference is that the target fluids used by SBC are liquid scintillators, so that nuclear recoils simultaneously create scintillation light and nucleate a bubble, as shown in Fig.~\ref{fig:xebc}, while electron recoils create light only.  The scintillation signal gives a measure of the total energy deposited in the detector with up to $\sim$keV resolution;
by contrast, the acoustic signal accompanying bubble nucleation measured by PICO gives $\sim$MeV energy resolution \cite{Amole:2015pla,Amole:2015lsj}.

Liquid-noble bubble chambers are also able to run at much lower thresholds (higher superheat) than Freon bubble chambers while remaining blind to electron recoils.  In Freon chambers, sensitivity to electron recoils turns on sharply at a Seitz threshold of a few keV, setting a lower limit on useful thresholds for dark matter searches.  Liquid-noble bubble chambers have fundamentally suppressed sensitivity to electron recoils, with ER bubble nucleation rates several orders of magnitude below that predicted by the models applied to Freon chambers (see Fig.~\ref{fig:xebc_ERdata}).  To our knowledge, bubble nucleation by electron recoils has never been observed in pure xenon~\cite{Brown:1956pgi}, and has been observed only at extreme superheat (thresholds of $\mathcal{O}(10)$\,eV) in pure liquid argon (LAr) \cite{HARIGEL1981363,PELLETT1963373}.

The absence of bubble nucleation by electron recoils in superheated noble liquids is a consequence of the lack of molecular degrees of freedom for recoiling electrons to excite.  
Without energy loss to molecular vibrational modes, the only way particle interactions can locally heat the fluid is through the center-of-mass motion of individual atoms.  While this is achieved directly by the atom-atom collisions that are the dominant mode of energy loss for nuclear recoils (\emph{i.e.}, the Lindhard effect~\cite{Lindhard:1961zz,Lindhard:1963a}), the kinematics of elastic electron-atom collisions make this an extremely inefficient means of energy loss for electrons.  Instead, electron recoil energy is almost entirely carried away by scintillation light, IR radiation (including bremsstrahlung from electron-atom scattering), and slowly or incompletely recombining ionization.  This was well known in the bubble chamber heyday, and liquid-xenon tracking bubble chambers explicitly mixed in molecular components (ethylene and propane) to allow bubble nucleation by delta ray electrons~\cite{bolozdynya2010emission}.

The thermodynamic threshold at which electron recoils may become a background in pure noble liquids is not yet known (see existing limits in Fig.~\ref{fig:xebc_ERdata}), but the threshold at which thermal fluctuations can spontaneously nucleate bulk bubbles is well studied \cite{SKRIPOV1979169}, and in liquid argon corresponds to $\sim$1 bubble per ton-year at a 40-eV Seitz threshold.  This target threshold sets the required pressure (1.4--25~bara) and temperature (90--130~K) ranges for the SBC-10LAr device described in Section~\ref{sec:design}.

\subsection{Results from the Prototype Xenon Bubble Chamber}
\begin{figure}[!tb]
    \centering
    \includegraphics[width=6.4in, trim = 0 0 0 0, clip=true]{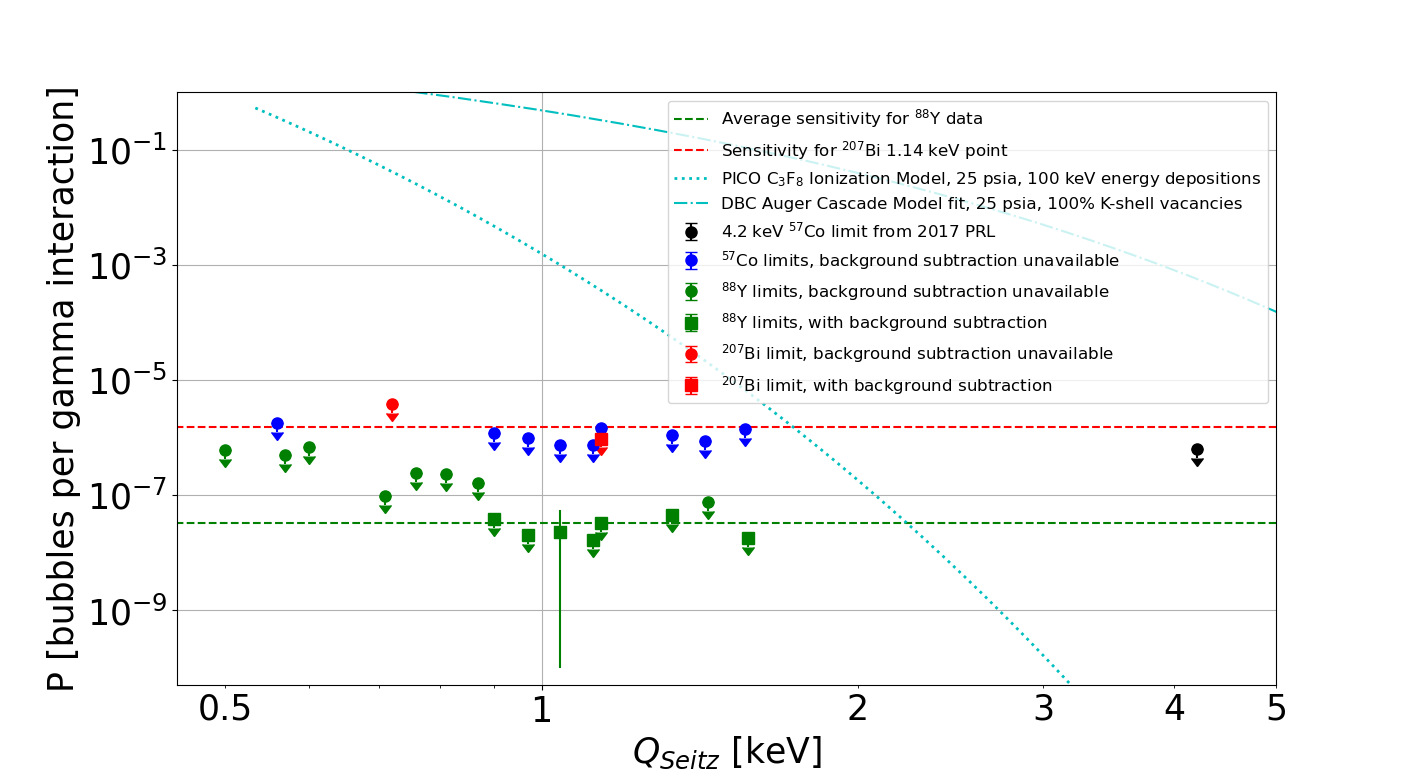}
    \caption{
    Electron recoil sensitivity of the prototype xenon bubble chamber as a function of superheat (Seitz threshold), taken from~\cite{BresslerPhD}.  No evidence for bubble nucleation by gamma sources is seen.  Dashed lines show the sensitivities of the measurements based on source strengths, background rates and exposures.  Square points show 90\% confidence intervals with background subtraction per \cite{Feldman:1997qc} (giving upper limits in all cases but one), and fall at the expected sensitivity for each source.  Round points are 90\% C.L. upper limits without background subtraction --- background data was not available at all operating conditions, particularly at the lowest thresholds.  The blue curves shows the gamma sensitivity expected for a molecular fluid with the same thermophysical properties as xenon, following the electron recoil bubble nucleation models for delta rays (dotted line, \cite{Amole:2019scf}) and Auger cascades following xenon K-shell capture (dot-dashed line, \cite{Bressler:2021ubm}).  
     }
    \label{fig:xebc_ERdata}
\end{figure}

\begin{figure}[!tb]
    \centering
    \includegraphics[height=2.7in, trim = 0 0 0 0, clip=true]{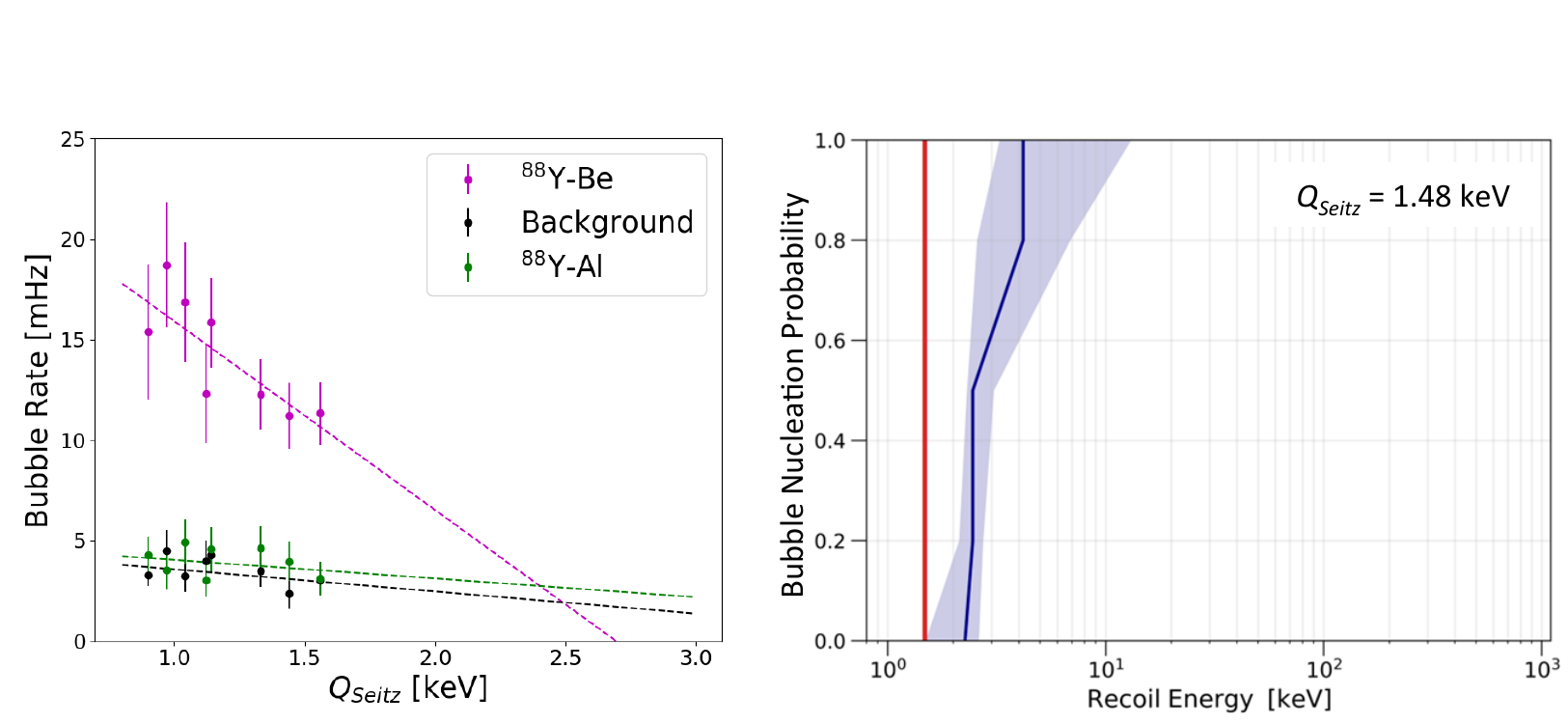}
    \caption{
    Nuclear recoil sensitivity of the prototype xenon bubble chamber --- both plots taken from~\cite{BresslerPhD}.  {\bf Left:} Chamber response to 152-keV neutrons from a $^{88}$Y-Be photo-neutron source.  The $^{88}$Y-Al points (gammas only) correspond to the $^{88}$Y points in Fig.~\ref{fig:xebc_ERdata}.  The extrapolated onset of sensitivity is at $Q\approx2.5$~keV for the 4.8-keV maximum recoil energy.  {\bf Right:} The probability of bubble nucleation as a function of nuclear recoil energy at a fixed Seitz threshold of 1.48~keV.  The efficiency curve is constrained by observed bubble rates when exposed to $^{252}$Cf, $^{88}$YBe, and $^{207}$BiBe sources, following the analysis strategy described in \cite{PICO:2022nyi}.  The solid blue line and shaded blue region indicate the best fit efficiency curve and 1-$\sigma$ allowed region, respectively, with the red line indicating the calculated Seitz threshold.  Both plots show a factor of $\sim$2 between the calculated Seitz threshold and nuclear recoil detection threshold, matching the threshold behavior seen in Freon bubble chambers \cite{Amole:2019fdf,Jin:2019thesis,PICO:2022nyi}. 
     }
    \label{fig:xebc_NRdata}
\end{figure}

The first detection of simultaneous scintillation and bubble nucleation by nuclear recoils was made in a 30-gram xenon bubble chamber at Northwestern University in 2016 \cite{Baxter:2017ozv} (see Fig.~\ref{fig:xebc}).  This chamber has since operated with xenon at thermodynamic thresholds down to 500~eV, limited only by the pressure rating of the device.  There is to-date no evidence for bubble nucleation by electron recoils in this device, giving the electron-recoil sensitivity limits shown in Fig.~\ref{fig:xebc_ERdata}.  The same device showed sensitivity to 152-keV neutrons from a $^{88}$Y-Be photo-neutron source \cite{Collar:2013ybe} (maximum xenon recoil energy of 4.8\,keV), confirming that the Seitz model remains an $\mathcal{O}(1)$ guidepost for nuclear recoil detection thresholds in noble liquids at thresholds down to at least 1~keV (see Fig.~\ref{fig:xebc_NRdata}).


\subsection{Strategies for Quasi-background-free Operation}
\label{ssec:bkgs}
For a dark matter search, the thermodynamic state of the bubble chamber will always be chosen such that the bubble nucleation rate from electron recoils (\emph{e.g.}~the Bq/kg of $^{39}$Ar beta-decays present in natural argon) is much less than one bubble per live-year.  There are, however, other backgrounds that do nucleate bubbles.  These are explained below, each with the corresponding mitigation employed to reach the goal of quasi-background-free operation.

\begin{figure}[!tb]
    \centering
    \includegraphics[width=6.4in, trim = 90 0 90 30, clip=true]{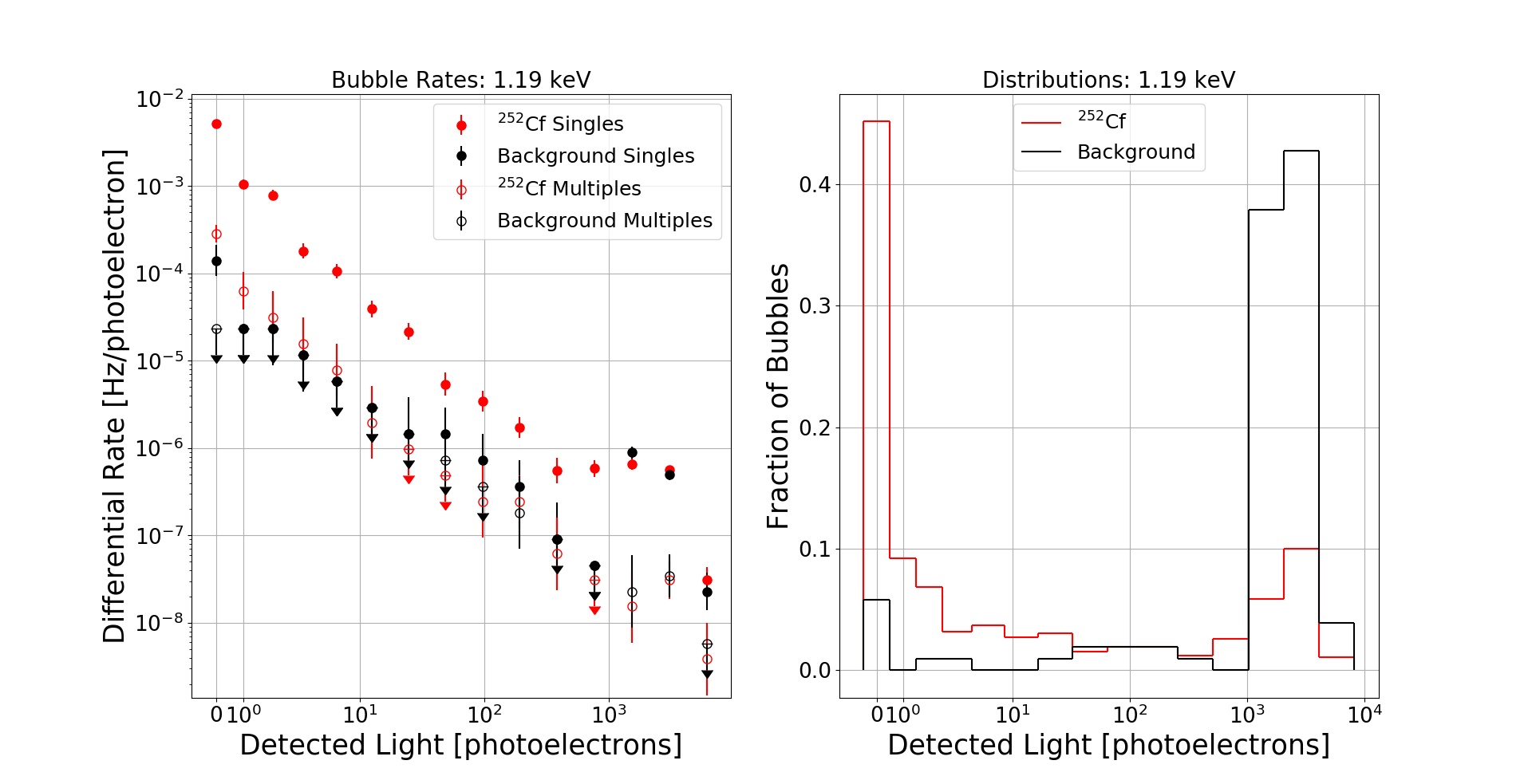}
    \caption{
    Scintillation response for a mock ``signal'' (in this case fast neutrons from a $^{252}$Cf source) and background in the prototype xenon bubble chamber, taken from~\cite{BresslerPhD}.  \textbf{Left:} The absolute rate, broken down by bubble multiplicity.  \textbf{Right:} The same data with unit normalization for both signal and background.  The background rate in this chamber consists primarily of high-energy events, including alpha-decays and through-going muons that nucleate a bubble by Coulomb scattering off a xenon nucleus, both of which saturate the scintillation detection system.  The SBC-LAr10 chamber will detect 1 scintillation photon per $\sim$5~keV nuclear recoil energy, or roughly $\times$10 the photon detection efficiency of the data shown here.
     }
    \label{fig:xebc_scintdata}
\end{figure}

Bulk alpha decays in the target fluid (\emph{e.g.}$^{222}$Rn, $^{220}$Rn, and the subsequent decay chains) nucleate a single bubble that is visually indistinguishable from those created by low-energy nuclear recoils, despite depositing $\sim$10$^4$ times more energy in the target.  Such bubbles are trivially identified by the accompanying scintillation signal, as was shown even with the rudimentary scintillation detection employed in the xenon prototype (see Fig.~\ref{fig:xebc_scintdata}).

Surface alpha decays may deposit much less energy in the target fluid, \emph{e.g.} giving only the tail end of a $^{206}$Pb recoil following $^{210}$Po decay, while still nucleating a bubble.  When this occurs at the chamber walls, these events are easily tagged by mm-resolution stereoscopic bubble imaging.  Standard low-background cleanliness protocols are necessary and sufficient to control the rate of such events from suspended particulate in the bulk target fluid.

Fast neutrons, produced by fission and $(\alpha,n)$ reactions in detector materials and by cosmic rays in the surrounding rock, are reduced primarily through shielding and careful screening of detector materials.  Any residual fast neutron flux can be measured and rejected by (1) tagging the multi-bubble events typical of a neutron's $\sim$10-cm mean free path, (2) using the scintillation channel to reject nuclear recoils greater than $\sim$5~keV (target scintillation detection threshold in SBC-LAr10), and (3) rejecting events with coincident scintillation in surrounding materials.  The last of these methods will become increasingly important as detectors grow in size, and may include scintillating thermal/hydraulic bath fluid (already in place in SBC-LAr10, see Section~\ref{sec:design}) and gadolinium-doped materials generating large neutron-capture gamma cascades to increase neutron-tagging efficiency.

Thermal neutrons can also nucleate bubbles in the target fluid through $(n,\gamma)$ (neutron capture) reactions in the target itself, where the capturing nucleus recoils against the high-energy (several-MeV) photons emitted.  As in the case of fast neutrons, this background is reduced primarily through shielding, with any residual rate identified by the scintillation produced by the correlated gammas interacting in the scintillating target and surrounding sensitive media.

In a similar fashion, external high-energy photons (1.5~MeV and above) can nucleate bubbles by generating nuclear recoils through Thomson scattering~\cite{Robinson:2016imi} --- the photon equivalent of the CE$\nu$NS signal.  These photons must be considered in shield design, and their flux can be monitored in-situ through the scintillation produced by their much more common electronic interactions.

Finally, if a positive signal inconsistent with the above backgrounds is observed --- \emph{i.e.} without an associated multi-bubble or scintillation-tagged event rate ---  several generic checks can be performed to rule out possible detector pathologies.  The NR bubble nucleation threshold can be varied to build a recoil energy spectrum, and moreover can be varied both by scanning in pressure or temperature to check for a consistent NR-like nucleation threshold.  A more powerful check, however, is to swap target fluid entirely --- all current SBC chambers are designed to run with either LAr or LXe targets (with the additional possibility of N$_2$ and CF$_4$)  granting a unique capability for self-verification of any future dark matter signal.

\clearpage
\section{Design and Construction of SBC-LAr10}\label{sec:design}

The SBC collaboration's first physics-scale detector, SBC-LAr10, is currently being commissioned at the Fermi National Accelerator Laboratory.  The primary purpose of this device is to characterize the low-threshold performance of the liquid-noble bubble chamber, validating the physics reach of the technique through the calibration campaign described in Section~\ref{sec:calibration}.  A functionally-identical device with a focus on radiopure construction will be deployed at SNOLAB for SBC's first dark matter search.  These chambers have been designed to meet the following specifications:

\begin{center}
\begin{tabular}{r|l}
\multicolumn{2}{c}{\bf SBC-LAr10 Design Goals} \\
\hline
\hline
    Target Volume &  10~L (10~kg LAr @ 130~K)\\
\hline
Target Fluid & Xe-doped Ar, \\
 & with options for pure Ar, Xe, N$_2$, and CF$_4$ \\
\hline
    Achievable Superheat & 40 eV (LAr @ 1.4~bara, 130 K)\\
\hline
Thermodynamic Regulation & $\pm$0.5~K, $\pm$0.1~bar ($\pm$5~eV Seitz Threshold)\\
\hline
Scintillation Detection & 1 photon per 5 keV NR in Xe-doped Argon \\
 & ($g_1\approx0.02$) \\
\hline
Bubble Imaging & 100~fps, mm-resolution stereoscopic imaging \\
\hline
Acoustic Reconstruction & time-of-nucleation reconstructed to \\
& $\pm$25~$\upmu$s resolution  \\
\hline
Zero-scintillation single-bubble rate & 1 background event per live year \\
(SNOLAB chamber)
\end{tabular}
\end{center}

\begin{figure}[bt]
\begin{center}
    \includegraphics[height=3in, trim=0 0 0 0, clip=true]{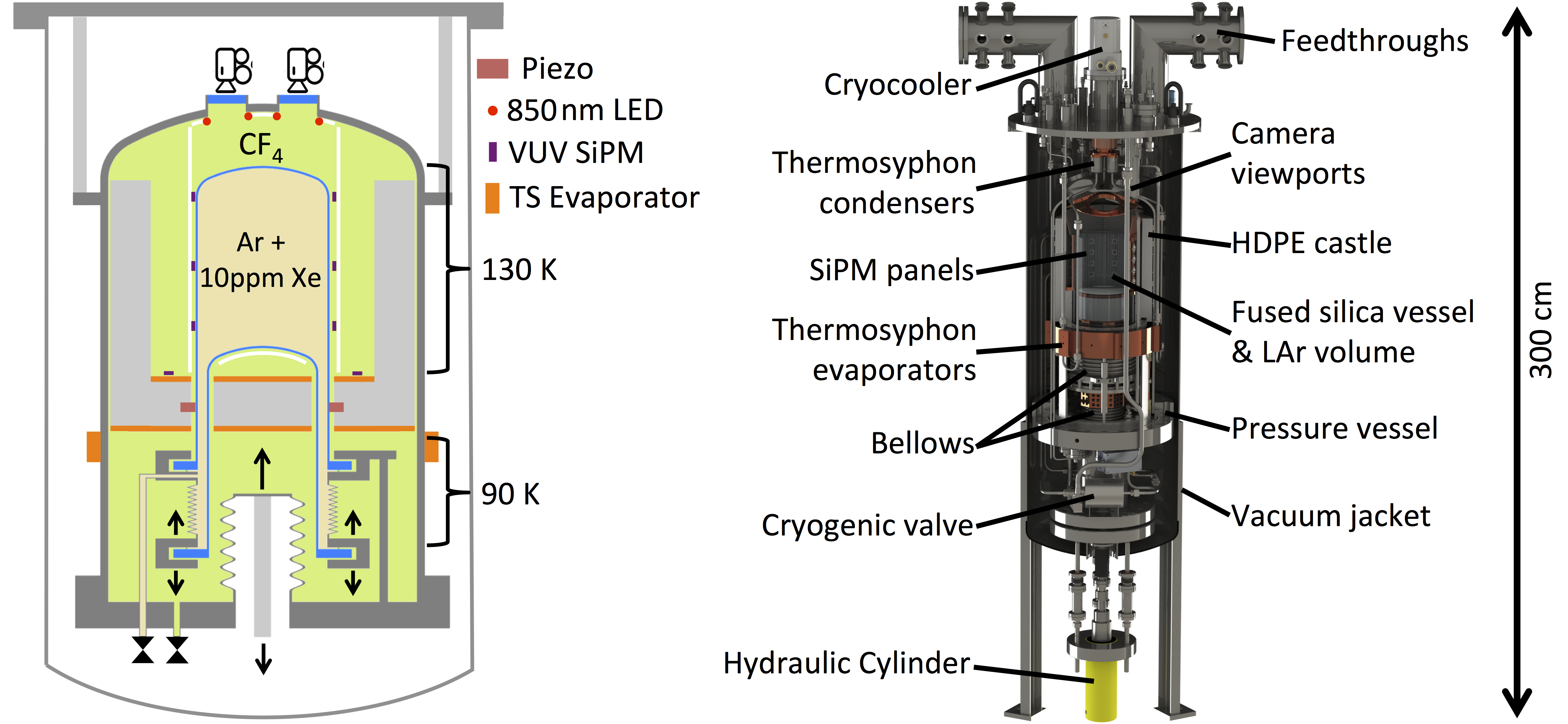}
\caption{Schematic (left) and annotated solid model (right) of the SBC-LAr10 detector.}
\label{fig:sbc}
\end{center}
\end{figure}

Figure~\ref{fig:sbc} shows the schematic and mechanical layout of the SBC-LAr10 detector.  The remainder of this section details the technical design of this detector, driven by the design goals listed above.

\subsection{Sites and Shielding}

As with all low-background detectors, SBC-LAr10 will require shielding in order to operate, however the requirements are very different for the two iterations of the detector.  Calibration of the detector will be a relatively low-rate process ($\sim$1,000 events per day, already much lower than the cosmic-induced background rate on surface), however the full dark matter search necessitates a dramatically reduced rate achievable only through multiple levels of shielding.

The calibration-focused chamber will be deployed at Fermilab in the MINOS tunnel, approximately 100~m underground.  This will reduce the cosmic-ray muon rate by an amount sufficient to allow the low-energy calibration without additional shielding.  The dark matter search chamber will be deployed at SNOLAB in the ladder labs (the space allocated following SBC's successful Gateway 1 review at SNOLAB) to achieve a significant reduction in the muon-induced background rate.  The detector will be deployed in a shield consisting of 50~cm of water to reduce the rock-based neutron rate and 10~cm of copper to attenuate high-energy gamma rays that would lead to a Thomson scattering background.

\subsection{Thermo-mechanical design}
As in other bubble chamber dark matter detectors, the SBC-LAr10 chamber must hold the target volume in the desired superheated state with at at least $\pm$0.5~K and $\pm$0.1~bar precision (corresponding to a $\pm$5~eV Seitz threshold at the target operating condition), waiting for a bubble nucleation event.  When bubble nucleation does occur, the chamber is recompressed within a fraction of a second to drive the fluid volume back to an entirely liquid state.  Both the precision regulation and fast compression must be achieved without exposing the superheated liquid to rough surfaces that can lead to spurious bubble nucleation.  SBC-LAr10 adopts a buffer-free, dual-temperature-zone strategy to achieve this, as is also used by PICO-40L and PICO-500~\cite{Giroux:2021vpy} and has been implemented successfully in SBC and PICO test chambers~\cite{Bressler:2021ubm,Baxter:2017ozv,BresslerPhD,Bressler:2019xhk, broerman2022new}.

\subsubsection{Pressure Control}
\begin{figure}[bt]
    \centering
    \includegraphics[height=10 cm, trim=15cm 25cm 20cm 5cm, clip=true]{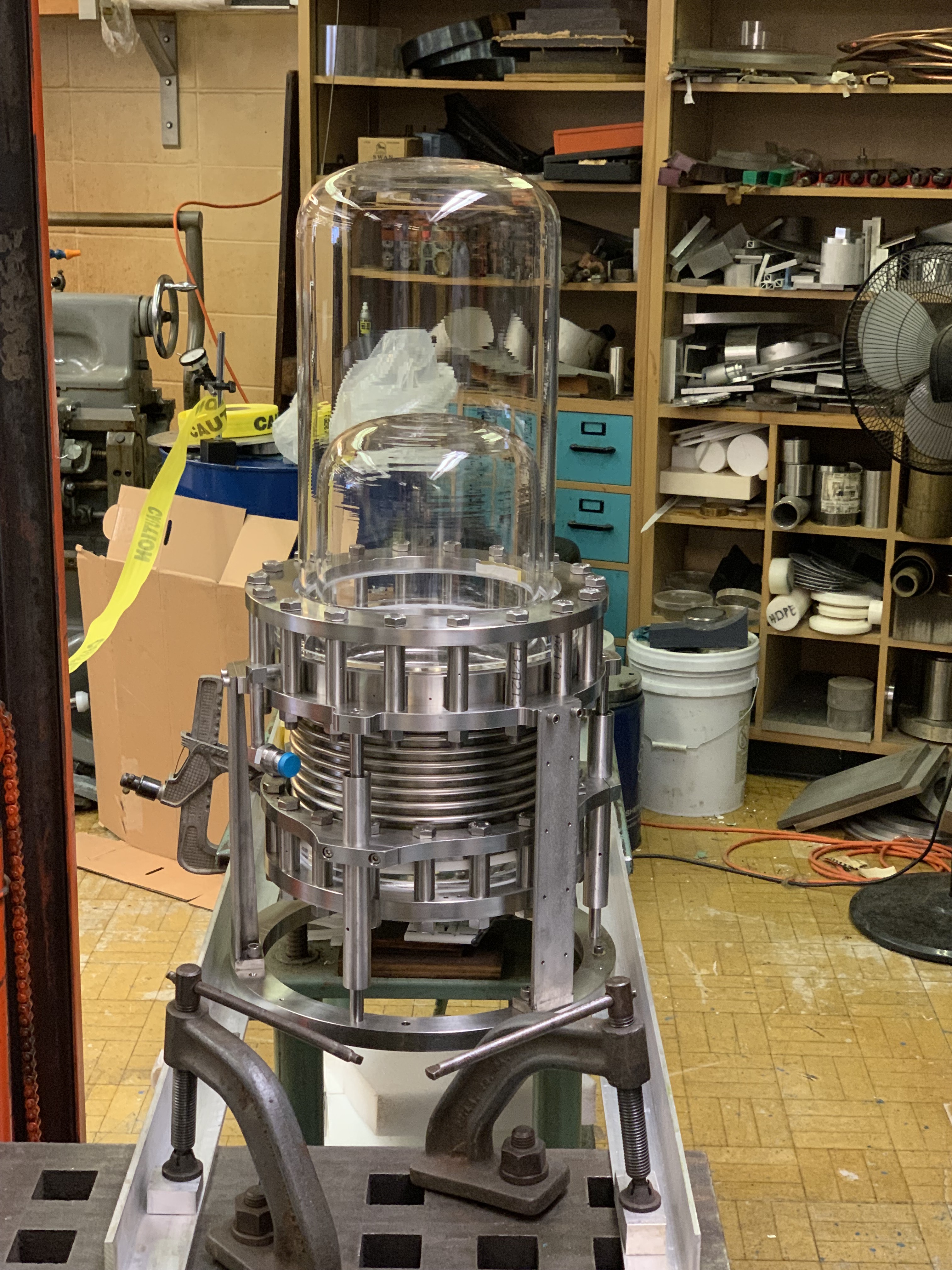} \includegraphics[height= 10cm, clip=true]{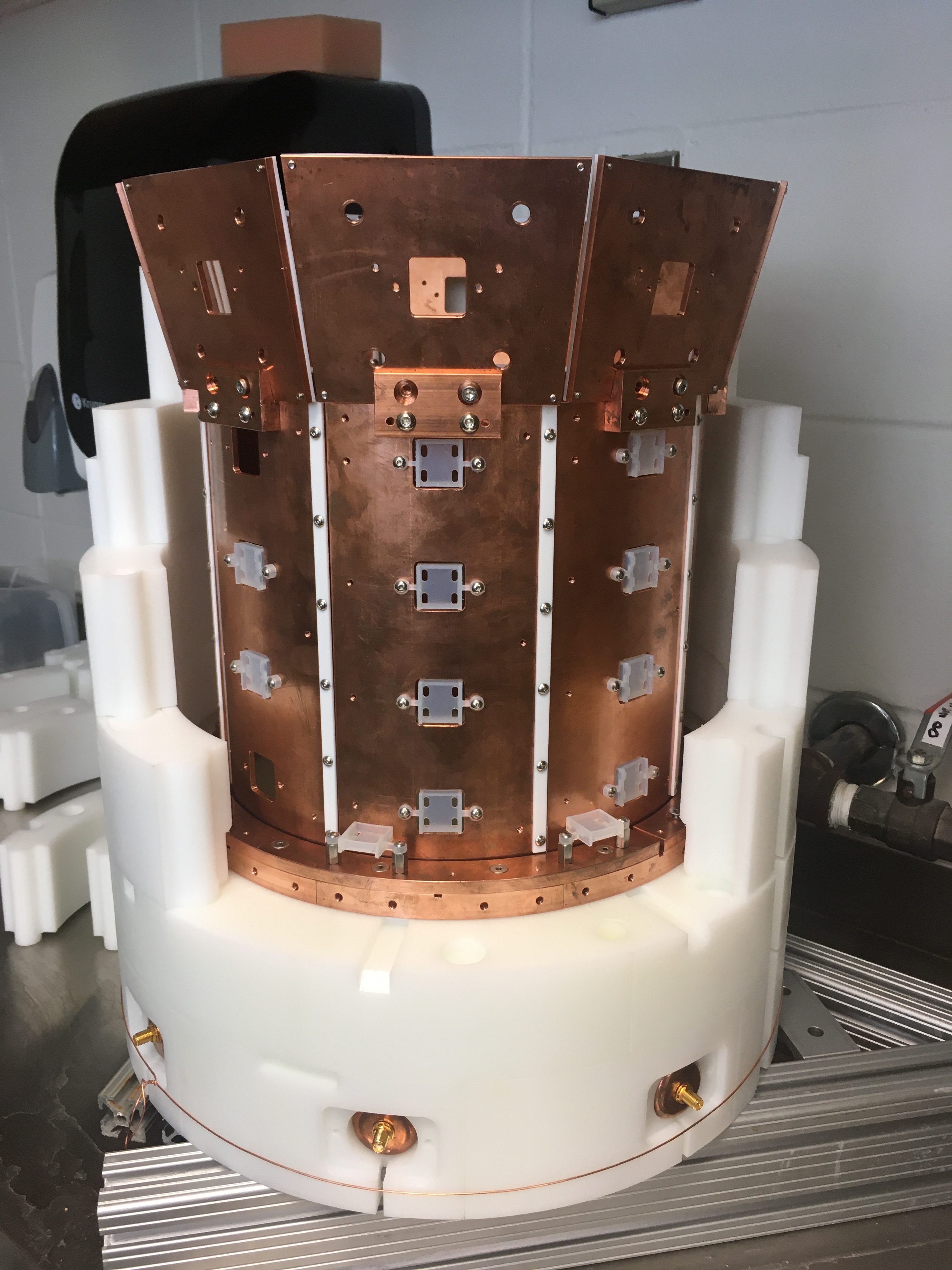}
    \caption{{\bf Left:} Test of the SBC-LAr10 ``inner assembly'' comprising the inner and outer fused silica vessels, stainless steel bellows, bellows guides and support structure, and spring-energized PTFE silica-to-metal seals. {\bf Right:} The partially-assembled HDPE ``castle'' isolating the warm (superheated) target region from the cold (stable) region, enclosing the copper SiPM support structure that surrounds the target volume.  At the bottom of the castle, three of the eight piezoelectric acoustic sensors are visible.}
    \label{fig:innerassembly}
\end{figure}

To avoid surface nucleation, the superheated fluid target is contained in a radiopure, UV-transparent, fused silica vessel (Hereaus Suprasil 310), with a second concentric, smaller-radius vessel acting as a piston to control the pressure of the target fluid.  Both vessels are sealed to a stainless steel bellows with spring-energized polytetrafluoroethylene (PTFE) seals.  The bellows, seals, and connected stainless steel piping are held at a lower temperature than the target fluid, such that the fluid contacting them is a stable liquid even in the depressurized state, with a temperature gradient (130~K to 90~K for LAr) in the annular space between the fused silica vessels.  A test assembly of the complete vessel-and-bellows system is shown in Fig.~\ref{fig:innerassembly} (left).

Typical operating pressures in SBC-LAr10 are 1.4~bara and 25~bara in the superheated and compressed states, respectively, well above the $\sim$1.3~bar differential pressure rating of the silica vessels.  The space outside the vessels is therefore filled with a cryogenic hydraulic fluid that drives the inner vessel ``piston'' to equalize the pressure inside and outside the silica vessels.  This hydraulic fluid must be a stable liquid (not superheated) over the full operating pressure range --- for operation with a LAr target the cryogenic hydraulic fluid is CF$_4$, a condensed inert gas with similar freezing point but much higher boiling point than LAr~\cite{Greer:1969a}.

\begin{figure}[tb]
    \centering
    \includegraphics[width=1.0\textwidth, trim=0 0 0 0, clip=true]{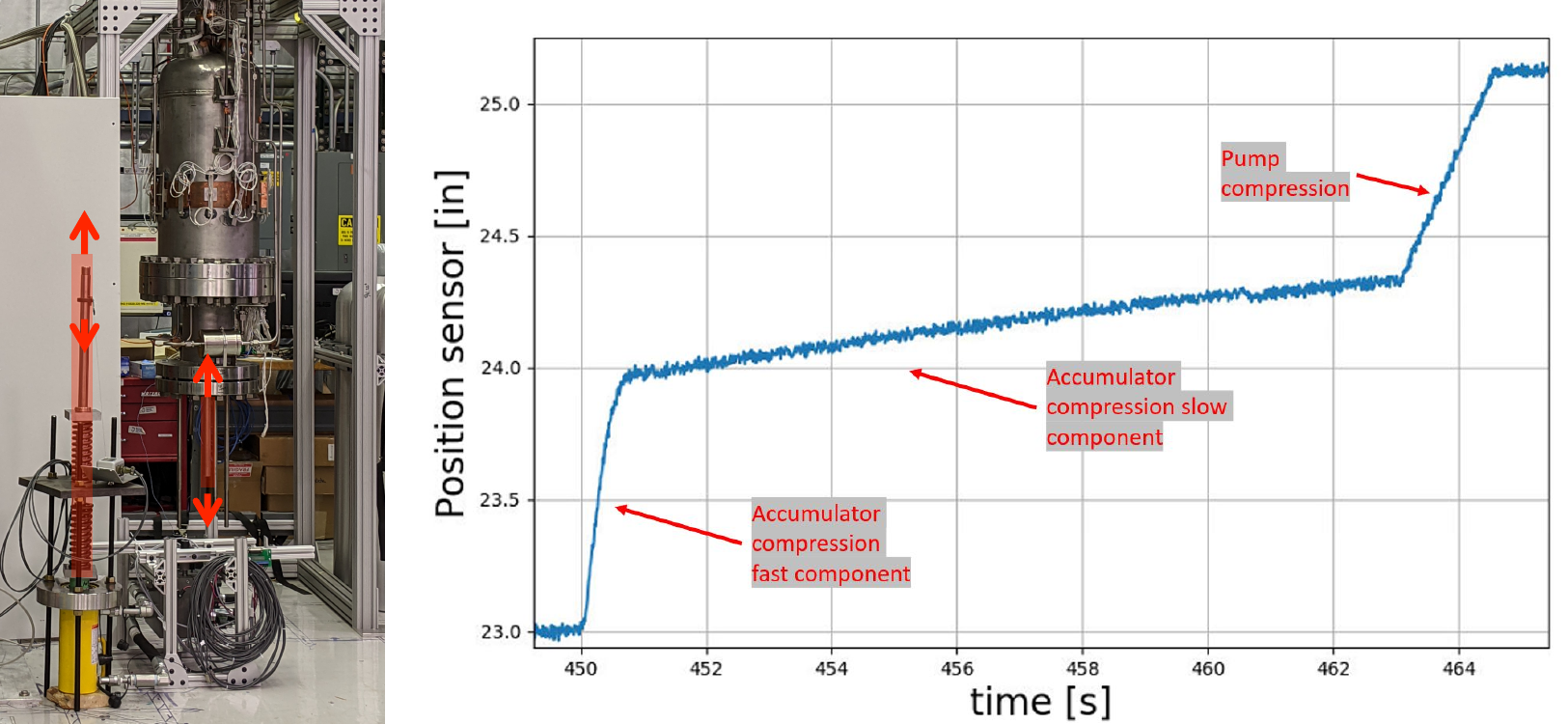}
    \caption{{\bf Left:} SBC-LAr10 pressure vessel beside the hydraulic test stand.  The hydraulic system was commissioned with the hydraulic cylinder (yellow) driving a pair of heavy springs.  In normal operation, the cylinder will connect to the bottom of the piston shaft seen at the base of the pressure vessel.  {\bf Right:} Sample ``fast compression'' in the hydraulic test stand, transitioning from a position-regulating mode to a high-pressure mode.  In normal operation, the cylinder will regulate chamber pressure rather than shaft position --- in the test stand the position regulation is limited by the resolution of the position transducer readout ($\sim$0.5~mm).  The initial sharp pressure rise driven by the accumulator is sufficient to halt bubble growth, while the subsequent slow rise driven by the hydraulic power unit drives the vapor back to the liquid state.  Figure taken from~\cite{BresslerPhD}.}
    \label{fig:hydraulictest}
\end{figure}

The cryogenic hydraulic fluid is contained in a stainless steel pressure vessel, and during bubble chamber operation both the hydraulic fluid and target fluid are isolated from the outside world by cryogenic valves immediately below the pressure vessel, creating a sealed space with fixed fluid mass.  The pressure of the cryogenic hydraulic fluid, and thus of the target fluid, is controlled by a bellows piston that is in turn driven by an external, room-temperature, double-acting hydraulic cylinder with 15-cm stroke and 55-ton capacity.  That cylinder is connected to a custom 2,000~psi hydraulic system with a 1.5-gpm hydraulic power unit, pneumatic bladder accumulator to drive fast compression, and servo valve for fine pressure control.  Figure~\ref{fig:hydraulictest} shows the completed hydraulic system and results from the first performance tests.

\subsubsection{Thermal Control}
To achieve the dual temperature zones necessary to avoid bubble nucleation on the stainless steel components and seals in the inner assembly, the SBC-LAr10 pressure vessel is cooled to the ``stable'' temperature (90~K for LAr) and an isolated region around the target fluid is heated to the superheated temperature (130~K for LAr at target threshold).  To achieve this gradient with an acceptable heat load, a high-density polyethylene (HDPE) ``castle'' is constructed around the target volume and annular target region, as shown partially assembled in Fig.~\ref{fig:innerassembly} (right).  Vertical tubes with custom check valves inside the castle connect the warm and cold regions at multiple locations to allow bi-directional fluid flow, preventing pressure build-up between the regions while restricting convection.  Flexible plastic ``wipers'' extend outward from the castle to the inner wall of the pressure vessel, further restricting convection.  A nichrome wire heater at the inside base of the castle provides the heat necessary to establish the warm temperature zone.

In the warm region the CF$_4$ liquid has low viscosity providing a highly-convecting, uniform thermal bath for the superheated target.  In the cold region the CF$_4$ is much more viscous and insulating, and inner-assembly metal components are thermally tied to the cold pressure vessel wall with spring-loaded copper fins.  Thin layers of liquid CF$_4$ serve as a thermal grease to complete the heat conduction path from the inner-assembly through the copper fins to the pressure vessel wall.

Heat is transferred from the pressure vessel wall to a copper band wrapping the exterior of the pressure vessel immediately below the bottom of the castle, seen in the left of Fig.~\ref{fig:hydraulictest}.  That band is in turn cooled by three closed-loop nitrogen thermosyphons~\cite{Bradley:2014lga}, each capable of carrying 100~W to a Cryomech AL300 cold-head immediately above the pressure vessel (see Fig.~\ref{fig:sbc}).  This 300-W Gifford-McMahon cryocooler finally rejects the heat to the outside world.

The pressure vessel, thermosyphons, and cold head are wrapped in multi-layer insulation and enclosed in a common vacuum jacket with penetrations for the compression system as well as fluid and electrical feedthroughs.  The total estimated heat load on the pressure vessel is 220~W, roughly half of which goes towards maintaining the thermal gradient between the warm and cold regions.

\subsubsection{Process Controls}
The complete SBC-LAr10 system includes approximately 125~channels of ``slow'' instrumentation --- pressure, temperature and position sensors, and the valves, heaters and pumps that control the chamber state.  These instruments, including the hydraulic pressure control system, the thermosyphon system, the target fluid (argon) and cryogenic hydraulic fluid (CF$_4$) fill and recovery systems, and the heaters and thermometry for temperature regulation are controlled by a Programmable Logic Controller, or PLC.  This PLC shares instrument readings in real time with a custom python-based SCADA (supervisory control and data acquisition, or ``slow-control'') system that provides a graphical user interface, records a continuous history of all instrument readings, and broadcasts alarms if detector parameters exceed allowed ranges.

The PLC also contains logic to regulate and manipulate the chamber state.  This logic is organized in three categories:  control loops, interlocks, and automated procedures. Control loops include PID-controlled heaters for temperature control and PID-control of the servo valve for pressure control. Interlocks are rudimentary if-then objects that can, for example, cut power to a heater or open a valve if a given temperature or pressure reading goes out of range.  Most interlocks are intended to protect detector hardware against unexpected conditions and operator error.  Automated procedures are sequences of steps that are either repeated too often to be human controlled (\emph{e.g.} the pressure cycling of the chamber for every event) or require a precision exceeding that achievable with manual controls (\emph{e.g.} the metered flow of N$_2$ gas into a thermosyphon circuit).

Finally, the PLC can also store 10-ms resolution traces of select instruments for asynchronous downloading, providing a more fine-grained view of chamber operation than can be achieved with the $\sim$1-second resolution provided by the real-time communication with the slow-control system.  This functionality produced the data shown in Fig.~\ref{fig:hydraulictest} (right), and is a key element of the data acquisition system described in the next section.

\subsection{Instrumentation and Data Acquisition}
Scintillating bubble chambers are somewhat unique in the variety of data, and the range of time scales over which data are taken, that are critical to event reconstruction.  From fast to slow, these include scintillation signals ($\mathcal{O}$(100)~ns), acoustic signals ($\mathcal{O}$(100)~$\upmu$s), bubble images ($\mathcal{O}$(10)~ms), and pressure and temperature histories ($\mathcal{O}$(0.01--1000)~s).  This section describes each of these signal pathways and the Run Control system that coordinates data acquisition between them.

\subsubsection{Event Structure}
All SBC-LAr10 data are organized by bubble chamber ``event'' as defined in Section~\ref{sec:concept} --- that is, one cycle from the pressurized state, to the superheated state, and back.  The transition from the superheated state back to the compressed state can be triggered in a number of ways, including detection of a bubble in the imaging system, detection of a pressure rise by the PLC, a random trigger generated by Run Control, or even an operator-generated manual trigger. In all cases this ``bubble chamber trigger'' is broadcast simultaneously to all data acquisition subsystems, allowing for offline synchronization of the different data streams.

The Run Control software initiates data acquisition in each data stream at the start of every event, and confirms the complete writing of data from each data stream after the bubble chamber trigger, closing the event.  For example, at the start of an event, Run Control instructs the PLC described in the previous section to begin recording 10-ms resolution data for later download.  Once all data subsystems are similarly primed, Run Control instructs the PLC to begin the pressure cycle.  On the bubble chamber trigger the PLC compresses the chamber.  Finally Run Control tells the PLC to halt acquisition and copies the PLC data to the event record.  The other data streams described below follow similar patterns.

\subsubsection{Bubble Imaging}
Three cameras continuously image the chamber, providing both a trigger on bubble nucleation and video record of the growing bubble(s) in the target volume.  SBC-LAr10 uses Arducam OV9281 1-Megapixel global shutter cameras, mounted in the vacuum space and viewing the chamber through three separate sapphire pressure windows at the top of the pressure vessel.  Three RaspberryPi~--~4B's (RPi's) mounted immediately outside the vacuum jacket capture images from each camera at 100 frames-per-second, storing images in a ring buffer.  The RPi also runs a simple motion detect algorithm, generating a bubble chamber trigger if a minimum number of pixels fluctuate by more than a set threshold between successive images.  When the bubble chamber trigger (which may or may not originate from the RPi) is received, the RPi takes a set number of post-trigger images and writes the images in the ring buffer to disk, typically capturing the $\sim$100~ms before and after bubble nucleation.

\begin{figure}[bt]
    \centering
    \includegraphics[width=1.0\textwidth, trim=0 45 0 30, clip=true]{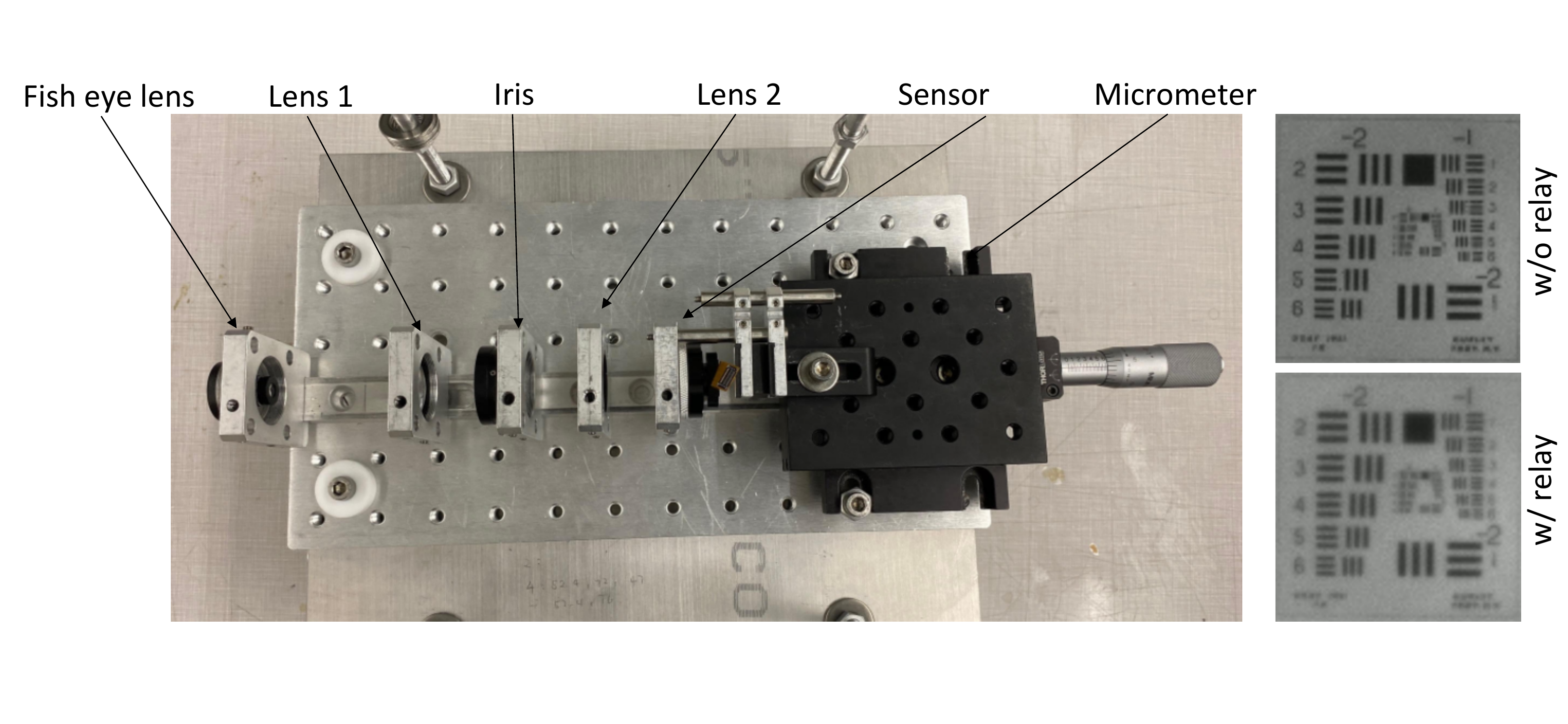}
    \caption{{\bf Left:} Relay lens test stand, transferring an image 17~cm from the objective fisheye lens (1.65-mm focal length) at the left end of the relay to the OV9281 sensor at the right. {\bf Right:} Sample zoomed-in images taken using the OV9281 Arducam with the objective lens mounted directly to the sensor (above) and transferred by the relay lens (below).}
    \label{fig:cameras}
\end{figure}

Tests with the Arducam OV9281 have shown it to be vacuum compatible but not compatible with cryogenic temperatures, making it necessary to separate the OV9281 sensor from the camera optics.  This is accomplished with a fused silica relay lens system, shown in the left of Fig.~\ref{fig:cameras}.  This system also serves to distance the camera sensor from the target volume, relaxing radiopurity constraints on the sensor in the SNOLAB chamber.  Tests are underway to verify that the relay lens systems maintains focus through thermal cycling of the chamber.

Illumination for bubble images is provided by three rings of 850-nm LEDs immediately inside the pressure vessel, providing a front-lit view of the PTFE-lined inner assembly.  The LEDs are pulsed in sync with the camera exposures (which are also synced between cameras), with a typical 10\% duty cycle to allow scintillation detection during the 90\% of the time with the chamber is dark.  A custom LED driver ensures stable illumination with sharp rise- and fall-times in the LED current.

Both the camera illumination and scintillation detection system (see next section) have been tested in a cryogenic CF$_4$ pressure test cell.  Both components function at the required temperatures and pressures, but the tests revealed a significant level of fluorescence in the cell, with an elevated single-photon detection rate lasting many seconds after each illumination flash.  If the same fluorescent behavior is seen in SBC-LAr10, it is possible to operate the imaging system with illumination active only \emph{after} the bubble chamber trigger is received, sacrificing the video trigger in favor of a darker chamber for scintillation detection while preserving the post-trigger images necessary for event reconstruction.

\subsubsection{Scintillation Detection}
Vacuum ultra-violet scintillation photons are detected by silicon photo-multipliers (SiPMs) immediately outside the target volume (\emph{i.e.} immersed in the cryogenic hydraulic fluid). The copper mounts for the SiPMs are shown in the right of Fig.~\ref{fig:innerassembly}.  Because the 128-nm argon scintillation light is absorbed by the fused silica walls of the inner assembly, the argon target is doped with $\mathcal{O}$(10)~ppm xenon, shifting the scintillation wavelength to 175-nm~\cite{Akimov:2017ble,Wahl:2014vma,Neumeier:2015ori,Peiffer:2008zz,Segreto:2020qks,Bernard:2022zyf} where both the fused silica and CF$_4$ hydraulic fluid have long absorption lengths.  All surfaces visible to the target volume except for the SiPMs, camera viewports, and annular region between vessels are covered with reflective PTFE, giving an estimated light collection efficiency of $\sim$10\% (dependent on the reflectivity of PTFE in liquid CF$_4$ and on the CF$_4$ absorption length, neither of which is yet well measured), which combined with the 20--25\% quantum efficiency of the VUV SiPMs achieves the target 2\% photon detection efficiency.

\begin{figure}[bt]
    \centering
    \includegraphics[width=1.0\textwidth, trim=0 20 0 20, clip=true]{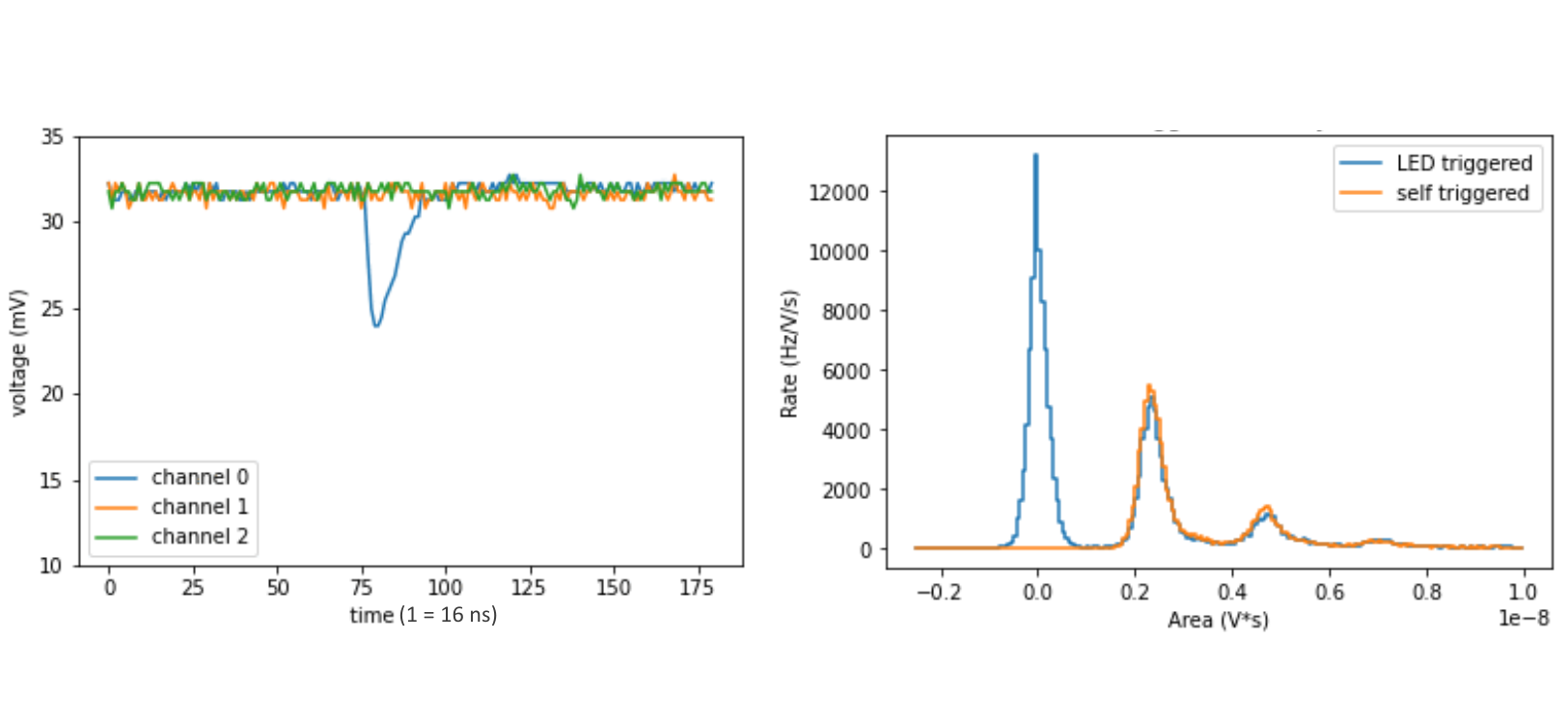}
    \caption{{\bf Left:} Three simultaneous traces from separate SiPMs in a Hamamatsu VUV4 Quad package, operating inside a pressurized, cryogenic liquid CF$_4$ test cell.  One SiPM trace shows a single photon (or single pixel avalanche).  {\bf Right:} Histogram of LED calibration data showing both the clear separation between the noise and single-photon peaks as well as the CAEN's ability to efficiently self-trigger on single photons.  Separation between discrete photon peaks is clear in larger signals as well, persisting up to at least 10 photons.}
    \label{fig:sipmdata}
\end{figure}

SBC-LAr10 at Fermilab uses Hamamatsu VUV4 SiPMs.  Gamma screening of these sensors reveals them to be far too high in U/Th content for a dark matter search, given their location immediately outside the target volume, so SBC has joined nEXO in procuring low-background SiPMs from FBK for the SNOLAB SBC-LAr10 clone.  In both cases, the SiPMs are powered by custom bias boards designed and produced at TRIUMF.  This package also provides amplification of the SiPM signals, which are then passed to a CAEN 62.5-MS/s multichannel digitizer.  The CAEN is configured to self-trigger when any input channel exceeds a set threshold (with $>$95\% efficiency for triggering on single photons, see Fig.~\ref{fig:sipmdata}), recording thousands of $\sim$1-$\upmu$s waveforms over the course of an event.  Both the individual waveform triggers and the bubble chamber trigger are timestamped by the CAEN, allowing for later synchronization with other data streams.

Figure~\ref{fig:sipmdata} shows the signal-to-noise observed from a Hamamatsu VUV4 SiPM immersed in a pressurized, cryogenic CF$_4$ test cell, using the full SiPM electronics chain.  These tests also indicated that liquid CF$_4$ generates $\mathcal{O}$(100) optical-wavelength scintillation photons per MeV deposited by external gamma sources, and 16 of the 48 SiPM mounting locations shown in Fig.~\ref{fig:innerassembly} are positioned to collect this light, providing an additional scintillation veto for external backgrounds.  CF$_4$ scintillation has been previously studied in gaseous CF$_4$~\cite{Pansky:1994zh,Margato:2013gqa}, but this is to our knowledge the first use of scintillation from liquid CF$_4$.  A publication on liquid CF$_4$ scintillation is in progress.

\subsubsection{Acoustic Detection}

In previous low-background bubble chambers, analyzing the acoustic information from each event has been vital to the discrimination of events generated by nuclear recoils from those generated by alphas~\cite{Amole:2019fdf}.  This role has largely been supplanted with the addition of information from the scintillation, with acoustic data now used primarily to determine the precise time of bubble nucleation, with 25-$\upmu$s accuracy demonstrated in the xenon prototype~\cite{Baxter:2017ozv}.  The shockwave produced by the growing bubble and subsequent $\mathcal{O}$(10)-kHz ringing of the detector is sensed by a set of textured lead zirconate titanate (PZT) piezoelectric elements mounted in the HDPE castle and spring-loaded against the outer fused silica vessel (see right of Fig.~\ref{fig:innerassembly}).  These signals are then passed through a pre-amplifier (both cold and warm preamp options are being prepared, mounted inside and outside the vacuum jacket, respectively) and digitized at 1.25~MS/s over a window extending $\sim$100~ms before and after the bubble chamber trigger.  A thorough discussion of these sensors, both in terms of constituents and use, can be found in \cite{Amole:2015pla}.

\subsection{Offline Analysis}

Interpretation of the raw data occurs in multiple modules, with the information from each module stitched together to form the full event. These modules define basic data quality cuts to catch excursions in pressure and temperature control, identify bubbles in camera images, reconstruct the 3-D position of each bubble, determine the time of nucleation from acoustic information, and identify any scintillation signal coincident with bubble nucleation.  All scintillation signals, whether coincident with a bubble or not, are analyzed to build the background LAr scintillation and CF$_4$ veto spectra. 

The 2-D bubble positions in each image can be determined by identifying clusters pixels with significant frame-to-frame variation~\cite{Amole:2015pla}.  Each 2-D bubble position corresponds to a ray through the bubble chamber, and from multiple images the point of closest approach of multiple rays can be identified, giving both the 3-D bubble position and a goodness-of-fit (distance of closest approach).

A template fit to a spectrogram of the acoustic signal from each acoustic sensor gives the time the bubble shockwave reaches each sensor.  A spread of $\sim$100-$\upmu$s is expected due to the speed of sound in the target fluid and silica vessel walls, varying with bubble position.  The position dependence of the acoustic lag will be calibrated using bubbles with unambiguous coincident scintillation pulses (giving the true time of nucleation), generating a position-corrected, acoustically-reconstructed time of nucleation with a target precision of $\pm$25~$\upmu$s.

Once the bubble position and time of nucleation are determined, coincident scintillation in the target or hydraulic fluid can be identified. For a dark matter search with an energy region-of-interest of 0.1~keV to 10~keV, signal events are not expected to create scintillation light, and an event can be rejected as a dark matter candidate when coincident with any scintillation signal.  Conversely, events with expected associated scintillation, either from a high energy recoil or coincident gamma ray emission, can be tagged by the scintillation signal --- this can be useful for the calibration studies described in the following section.

\section{Calibration Strategy for SBC-LAr10}\label{sec:calibration} 

The immediate objective of the SBC-LAr10 chamber is the determination of background rejection as a function of the nuclear recoil bubble nucleation threshold in superheated LAr.  This calibration will be entirely data-driven, with no input from the thermodynamic model described in Section~\ref{sec:concept}.  

\subsection{Electron Recoil Discrimination Calibrations}
The SBC calibration strategy begins by identifying the target operating condition (pressure and temperature) of the chamber by observing the onset of bubble nucleation by electron recoils, determined by exposing the chamber to a strong (mCi) gamma source.  
External gamma sources will also be used to calibrate the scintillation light yield in both the target LAr volume and in the outer LCF$_4$.  These light yields do not play a major role in the nuclear recoil calibrations below, as most sub-keV nuclear recoils will have no detected scintillation light --- but they are of key interest to the background characterization and mitigation strategy for later rare event searches.  


\subsection{Nuclear Recoil Sensitivity Calibrations}
\label{sec:calanalysis}

The approach to nuclear recoil calibration in SBC differs significantly from other sub-keV nuclear recoil experiments.  
Because the nuclear recoil region of interest for low mass dark matter ($\sim$100\,eV--1\,keV) is below SBC's scintillation detection threshold ($\sim$5\,keV per photon detected), the chamber is effectively a threshold detector and calibrations must rely on integrated recoil energy spectra.  Furthermore, because the total nuclear recoil detection rate in the chamber is limited to $\mathcal{O}(10^3)$~events per day, high-energy neutron scattering experiments where much of the rate is far above threshold (including most tagged scattering experiments), are effectively unfeasible.  These difficulties are offset by the extreme insensitivity of the device to electron recoils, making sources with high rates of ER background, including photo-neutron sources \cite{Collar:2013ybe,PhysRevC.94.024613}, high-energy gamma sources for Thomson scattering~\cite{Robinson:2016imi}, and thermal neutron capture viable options to generate low-energy nuclear recoils~\cite{Fischer:2019qfr,DanielDurnfordMSthesis}.  Each of these calibration techniques will be utilized as described below.

\subsubsection{Photoneutron sources}

\begin{figure}[tbh!]
    \centering
    \includegraphics[height=1.7in,trim= 0 0 50 50 ,clip=true]{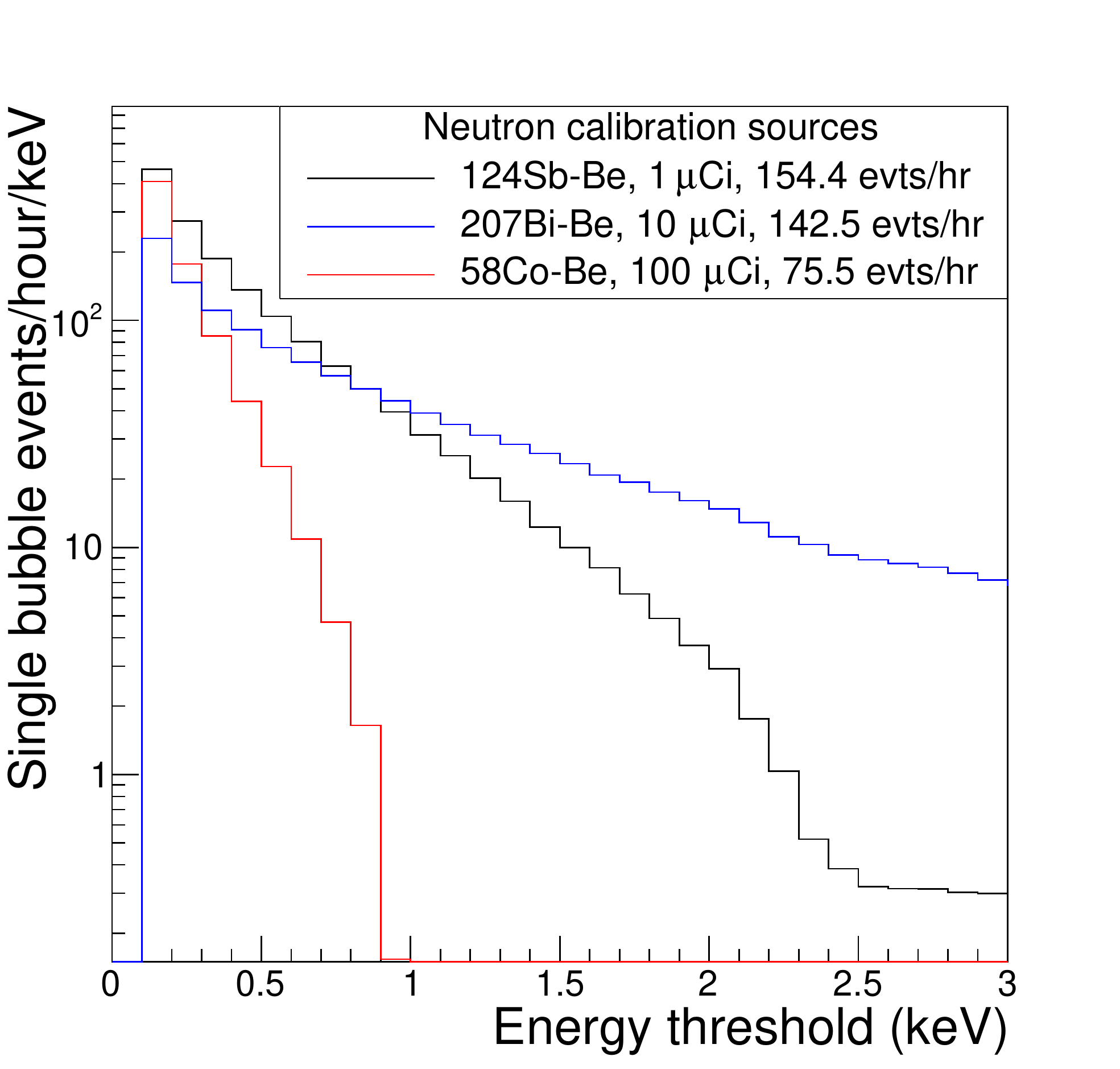}
    \includegraphics[height=1.7in,trim= 100 240 130 250,clip=true]{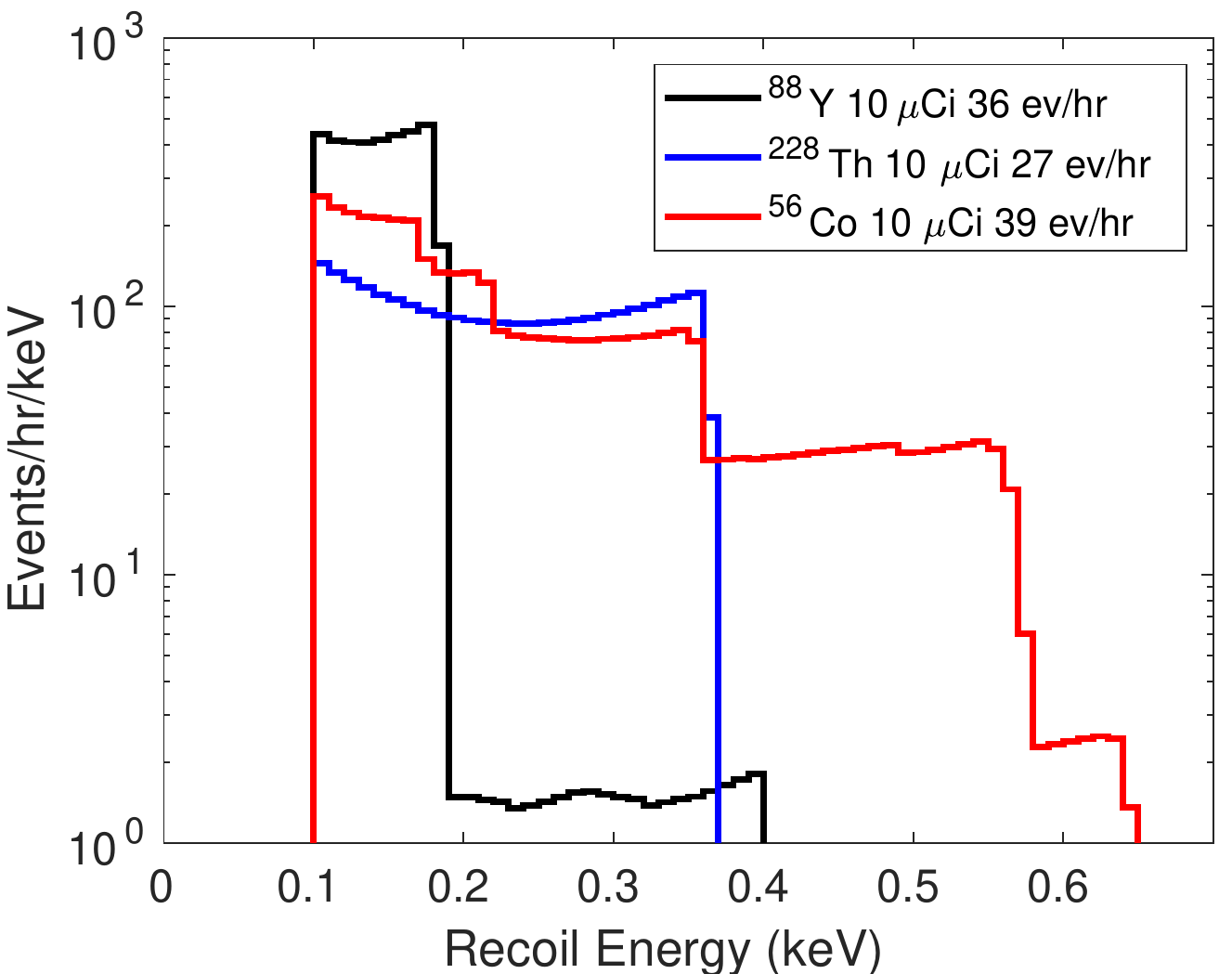}
    \includegraphics[height=1.7in, trim=0 0 0 0,clip=true]{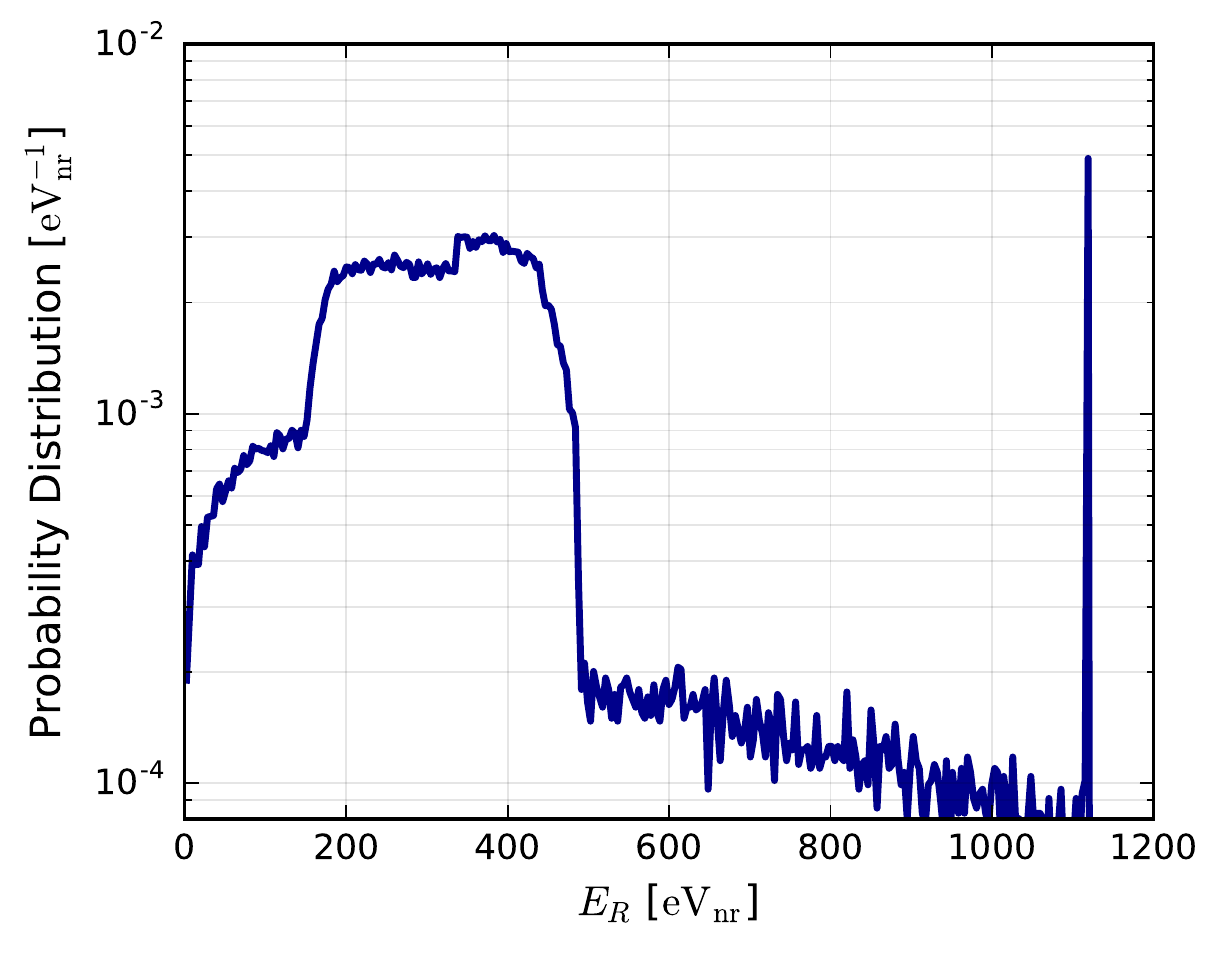}
    \caption{Simulated nuclear recoil spectra in the SBC-LAr10 chamber for photoneutron scattering (left), high energy photon-nucleus Thomson scattering (center), and thermal neutron capture  (right).  The capture nuclear recoil spectrum for natural argon is taken from \cite{DanielDurnfordMSthesis}.  
    }
    \label{fig:NRspectra}
\end{figure}

Once the gamma calibration program has identified the target operating temperature and pressure, the chamber will be exposed to a sequence of three photoneutron sources, each producing a different mono-energetic neutron, leading to different recoil energy spectra in the detector, as simulated in Fig.~\ref{fig:NRspectra}.  The different sources also lead to different bubble multiplicity in the detector, easily resolved with stereoscopic imaging.
The ability to measure bubble multiplicity in the chamber is crucial, as (1) ratios of rates at different multiplicities are insensitive to uncertainties on source strength, and (2) events with different bubble multiplicities sample different portions of the nuclear recoil spectrum, with lower recoil energies probed in high-multiplicity events.  
This calibration strategy was used successfully at few-keV thresholds by the PICO collaboration \cite{Amole:2019fdf,PICO:2022nyi}, and can be extended to $\sim$1-keV threshold with the photon-neutron source choices shown in Fig.~\ref{fig:NRspectra}.

\subsubsection{Photon-nucleus scattering}
A second nuclear recoil calibration technique, also taking advantage of the bubble chamber's insensitivity to electron recoils, is to look for photon-nucleus Thomson scattering by high-energy gamma rays.  For example, the 2.6-MeV gamma from $^{208}$Tl ($^{228}$Th decay product) will produce argon recoils up to 370~eV.  These sources do not generate multi-bubble events, but they do produce much sharper low-energy spectral features than the photoneutron sources (see Fig.~\ref{fig:NRspectra}).

\subsubsection{Thermal neutron capture}
Thermal neutron capture on $^{40}$Ar results in a 6.1-MeV gamma cascade with most of the energy typically released in a 4.7-MeV gamma ray \cite{Fischer:2019qfr,NESARAJA20161}, producing a $^{41}$Ar recoil spectrum peaked at $\sim$320~eV.  Thermal neutron capture on $^{36}$Ar is also significant despite its low abundance, giving a nearly mono-energetic 1.1-keV $^{37}$Ar recoil (see Fig.~\ref{fig:NRspectra}).  Neutron capture events may be tagged via the scintillation signal from the gamma cascade, which for $^{41}$Ar typically includes a characteristic 167-keV gamma from the end of the cascade.  

\subsubsection{Analysis techniques to constrain bubble nucleation efficiency}
Each of the measurements described above can be used to constrain the nuclear recoil nucleation efficiency curve, or the probability of bubble nucleation as a function of recoil energy at a fixed thermodynamic condition (see \emph{e.g.}~the right-side of Fig.~\ref{fig:xebc_NRdata}).  SBC and PICO collaborators have together developed a Markov-Chain Monte-Carlo-based method to combine constraints from diverse calibration sources with appropriate handling of systematic uncertainties \cite{PICO:2022nyi}, mitigating the different systematic uncertainties associated with each measurement.
SBC will use this method to achieve 20\% energy resolution on the nuclear recoil bubble nucleation threshold in SBC-LAr10, sufficient to determine the dark matter reach for the LAr bubble chamber technology.

\subsection{Extended calibration program}
As the first attempt at an ER-blind, sub-keV-threshold bubble chamber, it is entirely possible that SBC-LAr10 will not initially achieve the desired ER discrimination at low threshold.  In this eventuality, there are at least four pathways to follow, using the same SBC-LAr10 hardware, to improve low-threshold discrimination:  (1) Operation without (or with reduced) xenon doping, eliminating potential Auger-cascade backgrounds described in \cite{Amole:2019scf,Temples:2021jym,Bressler:2021ubm}; (2) Operation with pure LXe, potentially reducing prompt local heating associated with dimer formation along the electron track; (3) Operation with an applied $\mathcal{O}$(100)-V/cm electric field, reducing heating associated with electron-ion recombination; (4) Operation at reduced pressure and temperature, shown in \cite{Amole:2019scf} to reduce ER sensitivity at constant Seitz threshold.

On the other hand, if low-thresholds are immediately realized, extended calibrations will aim to achieve the 5\% threshold uncertainty needed for precision CE$\nu$NS physics, as described in~\cite{SBC:2021yal}.  Precision calibration will be enabled by a set of photon-nucleus Thomson scattering sources with recoil endpoints near the operating threshold~\cite{Lamb:2022aps}. Long term calibration plans with SBC-LAr10 will also include operation and threshold characterization in pure xenon as well as with N$_2$ and CF$_4$ targets, which can also be deployed in SBC-LAr10 and which, while molecular, are simple molecules with high-energy bonds and do produce scintillation light, indicating that low-threshold discrimination may be achievable in those fluids as well.

\section{Next Steps and Long Term Opportunities}\label{sec:future}
The future of SBC's liquid-noble bubble chamber program depends critically on the results of the work described in Sections~\ref{sec:design} and~\ref{sec:calibration}, and specifically on the achievable ER-blind NR detection threshold.  This section describes the steps to be taken after SBC-LAr10, assuming the 100-eV threshold motivated in Section~\ref{sec:concept} is achieved, in order to fulfill the science goals described in Section~\ref{sec:science}, as well as longer term opportunities that may become available as the liquid-noble bubble chamber technology matures.

\subsection{Dark Matter Next Step:  Solar CE$\nu$NS Floor at 1~GeV}
As shown in Fig.~\ref{fig:sbc_chamber_reach}, a 100-eV argon nuclear recoil detection threshold in a chamber that can discriminate electron recoils puts the solar CE$\nu$NS floor squarely in reach for an SBC device of similar scale to the PICO-500 detector, which will operate at SNOLAB concurrently with the radiopure SBC-LAr10~\cite{Giroux:2021vpy}.  As in every dark matter experiment, however, increases in exposure bring increased demands on background mitigation techniques, in this case for the bubble-nucleating backgrounds discussed in Section~\ref{ssec:bkgs}.  The primary strategy to suppress these backgrounds will be increased SiPM coverage coupled with additional scintillating material outside the superheated target, decreasing veto thresholds and giving additional opportunities to tag external backgrounds.

\subsection{Dark Matter Long Term Opportunities}
The neutrino fog is largely seen as the end of the road for searches for spin-independent (SI) dark matter-nucleus interactions.  However, by using low-A targets with large spin-dependent (SD) sensitivity and low SI sensitivity such as fluorine, nitrogen, and hydrogen, it is possible in principle to explore orders of magnitude more SD parameter space than is accessible to argon- and xenon-based experiments~\cite{Akerib:2022ort} with useful exposures up to a kton-year and beyond.  R\&D into the technology advances necessary for such a large scale detector, including new vessel materials~\cite{broerman2022new} and bubble imaging methods, would be performed in collaboration with PICO, as would any subsequent dark matter search. The background-rejection power of a scintillating superheated target such as CF$_4$ or N$_2$ may be a key element to a future deep exploration of the SD parameter space.  

\subsection{Reactor CE$\nu$NS Next Step:  First Detection}
\begin{figure}[tb]
    \centering
    \includegraphics[height=2.5in, trim=0 0 0 0, clip=true]{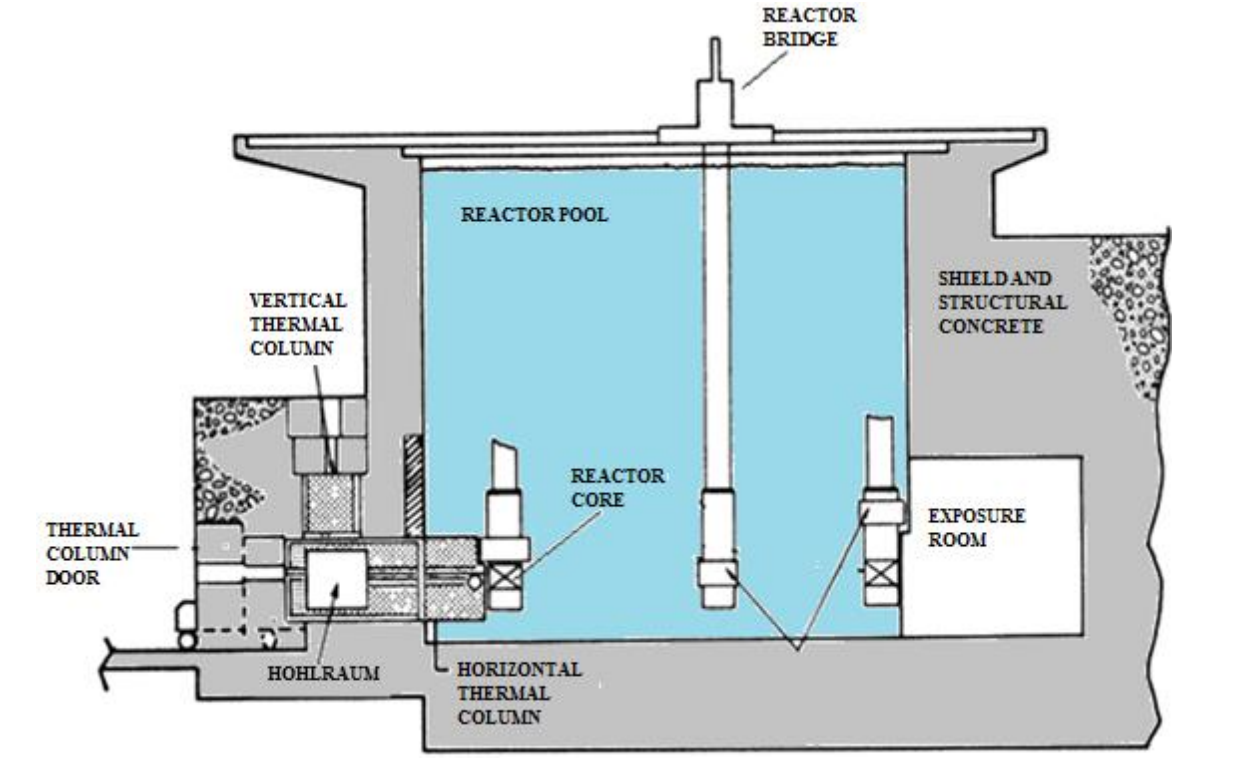}
    \caption{Cross section of the ININ TRIGA Mark III 1~MW$_{\rm{th}}$ reactor.  SBC collaborators are investigating the feasibility of operating an SBC device in the exposure hall, allowing baselines as short as 3~m from the movable reactor core.  Figure copied from~\cite{IAEA:ININ}.}
    \label{fig:ININ}
\end{figure}

The SBC-LAr10 chamber at 100-eV threshold is capable of detecting dozens (hundreds) of CE$\nu$NS events per day at a typical research (power) reactor site.  The challenge, once the threshold is achieved, is to locate a site with sufficient overburden and shielding (or space to build shielding) to suppress the rate of cosmic- and reactor-induced neutron events to below the CE$\nu$NS signal rate.  SBC collaborators are currently characterizing the neutron environment at the Instituto Nacional de Investigaciones Nucleares (ININ) TRIGA Mark III reactor near Mexico City~\cite{IAEA:ININ}, a 1~MW$_{\rm{th}}$ reactor with a movable core allowing as little as a 3~m baseline to an exposure room, with 3~m overburden to shield cosmic rays (see Fig.~\ref{fig:ININ}).

\subsection{Reactor CE$\nu$NS Long Term Opportunities}
The many possible reactor CE$\nu$NS programs described in Section~\ref{sec:science}, including measurements of neutrino interactions, sterile neutrino oscillation searches, and reactor monitoring, all benefit from the increased CE$\nu$NS rate that comes with a larger target mass.  This is at odds with the inherent rate limitations in bubble chambers --- even a 10-kg chamber requires significant shielding or overburden to reduce the ambient cosmic-induced bubble rate to levels below the $\sim$1,000 event/day maximum acquisition rate.  These events are mostly far above the CE$\nu$NS signal energy region of interest, but none-the-less inhibit the live time of the bubble chamber.  This live-time hit can be avoided by modularizing the detector so that only a portion of the target mass must be re-pressurized after each event.  Many strategies for this modularization are possible, but all would benefit from a more rugged design than the fused-silica vessels and sapphire viewports used in monolithic chambers, leading to similar underlying R\&D challenges to those in kton-scale dark matter detection: namely, identifying viable alternatives to silica vessels and optical bubble reconstruction.

\section{Acknowledgements}
This document was prepared by the Scintillating Bubble Chamber (SBC) collaboration using the resources of the Fermi National Accelerator Laboratory (Fermilab), a U.S. Department of Energy, Office of Science, HEP User Facility. Fermilab is managed by Fermi Research Alliance, LLC (FRA), acting under Contract No. DE-AC02-07CH11359.
The SBC collaboration also wishes to thank SNOLAB and its staff for support through underground space, logistical and technical services. SNOLAB operations are supported by the Canada Foundation for Innovation and the Province of Ontario Ministry of Research and Innovation, with underground access provided by Vale at the Creighton mine site.
The SBC-LAr10 chamber at Fermilab and its twin at SNOLAB have been supported primarily by the Fermilab Lab-Directed R\&D program and the Canada Foundation for Innovation (CFI), respectively, with additional support from the Natural Sciences and Engineering Research Council of Canada (NSERC), the DOE Office of Science Graduate Instrumentation Research Award fellowship, the Arthur B. McDonald Canadian Astroparticle Physics Research Institute, the projects CONACyT CB2017-2018/A1-S-8960, DGAPA UNAM grant PAPIIT IN108020, Fundaci\'on Marcos Moshinsky, the Indiana University South Bend Office of Research, the URA Visiting Scholar program, the Fermilab Cosmic Physics Center, and  DOE Office of Science grants DE-SC0015910, DE-SC0017815, and DE-SC0011702.


\bibliographystyle{jhep}
\bibliography{references.bib}

\end{document}